%% file: main.tex
\documentclass[sigconf]{acmart}

\graphicspath{{./figures/}}
\usepackage{enumitem}
\usepackage[tight,footnotesize]{subfigure} 
\usepackage{graphicx}
\usepackage{pifont}
\usepackage{longtable}
\usepackage[T1]{fontenc}
\usepackage[utf8]{inputenc}
\usepackage{letltxmacro}
\usepackage{xcolor,colortbl}
\usepackage{amssymb}
\usepackage[update,prepend]{epstopdf}
\usepackage{listings}
\usepackage{array}
\usepackage[ruled,vlined,linesnumbered]{algorithm2e}
\usepackage{tcolorbox}
\usepackage{multirow}
\usepackage{supertabular,booktabs} 
\usepackage{siunitx}
\usepackage{amsfonts}
\usepackage{fdsymbol}
\usepackage{makecell}
\usepackage{enumitem}
\usepackage [autostyle, english = american]{csquotes}
\usepackage{makecell}
\usepackage{tabstackengine}
\usepackage[title]{appendix}
\usepackage{multicol,lipsum}
\usepackage{multirow}

\definecolor{darkblue}{rgb}{0.0, 0.0, 0.55}

\newcommand{\etal}{\emph{et al.}}
\newcommand{\eg}{\emph{e.g.}}
\newcommand{\ie}{\emph{i.e.}}

\setcopyright{none} 
\acmConference[Anonymous Submission to ACM CCS 2019]{ACM Conference on Computer and Communications Security}{Due 15 May 2019}{London, TBD}
\acmYear{2020}

\setcopyright{none}
\begin{document}
\title{Measuring the Effectiveness of Privacy Policies \\for Voice Assistant Applications} 

\author{Song Liao, Christin Wilson, Long Cheng, Hongxin Hu, and Huixing Deng}
\affiliation{%
	\institution{School of Computing, Clemson University}
}
\begin{abstract}
Voice Assistants (VA) such as Amazon Alexa and Google Assistant are quickly and seamlessly integrating into people’s daily lives. 
The increased
reliance on VA services raises privacy concerns such as the leakage of private conversations and sensitive information. Privacy policies play an important role in addressing users' privacy concerns and informing them about the data collection, storage, and sharing practices.
VA platforms (both Amazon Alexa and Google Assistant) allow third-party developers to build new voice-apps and publish them to the app store. 
Voice-app developers are required to provide privacy policies to disclose their apps' data practices. However, little is known whether these privacy policies are informative and trustworthy or not on emerging VA platforms. On the other hand, many users invoke voice-apps through voice and thus there exists a usability challenge for users to access these privacy policies.

In this paper, we conduct the first large-scale data analytics to systematically measure the effectiveness of privacy policies provided by voice-app developers on two mainstream VA platforms. We seek to understand the quality and usability issues of privacy policies provided by developers in the current app stores. We analyzed 64,720 Amazon Alexa skills and 2,201 Google Assistant actions. Our work also includes a user study to understand users' perspectives on VA's privacy policies. Our findings reveal a worrisome reality of privacy policies in two mainstream voice-app stores, where there exists a substantial number of problematic privacy policies. Surprisingly, Google and Amazon even have official voice-apps violating their own requirements regarding the privacy policy.
\end{abstract}




\maketitle

\input{introduction}

\input{background}

\input{datacollection}

\input{policyanalysis}

\input{usersurvey}

\input{discussion}

\input{privacyintent}

\input{relatework}

\input{concludion}

\bibliographystyle{plain}
\bibliography{ref.bib}

\input{appendix}


\end{document}

%% file: introduction.tex
\section{Introduction}\label{Introduction}

Virtual Assistants (VA) such as Amazon Alexa and Google Assistant have been seamlessly integrated into our daily life. An estimated 3.25 billion digital voice assistants are being used around the world in 2019. The number is forecasted to reach 8 billion users by 2023 which is higher than the current world population~\cite{VA:2023:Number}. VA handles a wide range of queries that humans are posing, \eg, from ordering everyday items, managing bank accounts, controlling smart home devices to recommending clothing stores and new fashions. Despite the many convenient features, there is an increasing concern on privacy risks of VA users~\cite{Chung:2017:Computer, Maurice:2017:Law, Geeng:2019:WCI, MCLEAN201928, Ammari:2019:MSI, Alexander:2019, Malkin:2019:Privacy, Faysal}. 

Privacy and data protection laws are in place in most of the countries around the world to protect end users online. These compliance requirements are mostly satisfied by providing a transparent privacy policy by developers. Google was fined €50 million by a French data protection regulator after its privacy policy failed to comply with General Data Protection Regulation (EU GDPR). This fine was not for failing to provide a privacy policy but for not having a one that was good enough and failing to provide enough information to users~\cite{Google:2019:fined}. Researchers have shown that there are many discrepancies between mobile apps (\eg, Android apps) and their privacy policies~\cite{Slavin:ICSE:2016, Wang:2018:ICSE, Zimmeck:PETS:2019, SEC:Benjamin:2019}, which may be either because of careless preparation by benign developers or an intentional deception by malicious developers~\cite{Yu:2016:DSN}. Such inconsistencies could lead to public enforcement actions by the Federal Trade Commission~(FTC) or other regulatory agencies~\cite{zimmeckEtAlCompliance2017}. 
For example, FTC fined \$800,000 against Path (a mobile app operator) because of an incomplete data practice disclosure in its privacy policy~\cite{Path:2013:FTC}. In another case, Snapchat transmitted geolocation information from users of its Android app, despite the privacy policy states that it did not track such information. In 2014, FTC launched a formal investigation requesting Snapchat to
implement a comprehensive privacy program~\cite{Snapchat:2014:FTC}.


VA platforms allow third-party developers to build new voice-apps (which are called {\em skills} on Amazon platform and {\em actions} on Google platform, respectively) and publish them to app stores. In order to comply with privacy regulations (such as COPPA~\cite{Security:2019:COPPA}) and protect consumers' privacy, voice-app developers are required to provide privacy policies and notify users of their apps' data practices. Typically, a proper privacy policy is a document that should have answers to a minimum of three important questions~\cite{PrivacyPolicyGuidance}: 1) What information is being collected? 2) How this information is being used? and 3) What information is being shared? Third-party skills and actions are in very high number in the respective stores. Developers are required to provide privacy policies for their voice-apps. Policies could be diverse and poorly written, which results in more users ignoring the privacy policy and choosing to not read it. This also leads to users using a privacy-sensitive service without having a proper understanding of the data that is being collected from them and what the developer will do with it. On the other hand, the feature that makes VA devices like Amazon Echo and Google Assistant interesting is the ability to control them over the voice without the need of physically accessing them. Despite the convenience, it poses challenges on effective privacy notices to enable users to make informed privacy decisions. The privacy policy may be missing completely in the conversational interface unless users read it over the smartphone app or through the web.







In this work, we mainly investigate the following three research questions (RQs):
\vspace{-3pt}
\begin{itemize}
	\item RQ1: What is the overall quality of privacy policies provided by voice-app developers in different VA platforms? Do they provide informative and meaningful privacy policies as required by VA platforms?  
	\item RQ2: For a seemingly well-written privacy policy that contains vital information regarding the service provided to users, can we trust it or not? Can we detect inconsistent privacy policies of voice-apps?
	\item RQ3: What are VA users' perspectives on privacy policies of voice-apps? What is possibly a better usability choice for VA users to make informed privacy decisions? 
\end{itemize}


We conduct the first empirical analysis to measure the effectiveness of privacy policies provided by voice-app developers on both Amazon Alexa and Google Assistant platforms. Such an effort has not previously been reported. The major contributions and findings are summarized as follow\footnote{Accompanying materials of this work including the dataset, empirical evidences for inconsistent privacy policies, and tools are available at https://github.com/voice-assistant-research/voice-assistant.}.
\begin{itemize}
	\item We analyze 64,720 Amazon Alexa skills and 2,201 Google Assistant actions. We first check whether they have a privacy policy. For the 17,952 skills and 1,967 actions that have one, unfortunately, we find there are many voice-apps in app stores with incorrect privacy policy URLs or broken links.
	Surprisingly, Google and Amazon even have official voice-apps violating their own requirements regarding the privacy policy. 
	\item We further analyze the privacy policy content to identify potential inconsistencies between policies and voice-apps. We develop a Natural Language Processing (NLP)-based approach to capture data practices from privacy policies. We then compare the data practices of a privacy policy against the app's description. We find there are privacy policies that are inconsistent with the corresponding skill descriptions. We also find skills which are supposed to have a privacy policy but do not provide one.
	\item We conduct a user study with 91 participants to understand users' perspectives on VA's privacy policies, using the Amazon Mechanical Turk crowdsourcing platform. We also discuss solutions to improve the usability of privacy notices to VA users. 
\end{itemize}

%% file: background.tex
\section{Background and Challenges}\label{background}


\subsection{Voice-app and privacy policy} 

{\bf Voice-app listing on the store.} 
We mainly focus on two mainstream VA platforms, \ie, Amazon Alexa and Google Assistant, both with conceptually similar architectures. These platforms allow third-party developers to publish their own voice-apps on VA stores. A voice-app's introduction page that is shared by the developer on the store contains the app name, a detailed description, the category it belongs to, developer information, user rating and reviews, privacy policy link, and example voice commands which can be viewed by end users. The source code is not included in the submission and therefore is not available either to the certification teams of VA platforms or to end users. Users who enable a skill/action through the voice-app store may make their decisions based on the description. It explains the functionality and behavior of the voice-app and what the user can expect from it. Some developers also mention the data that is required from users (\ie, data practices) in the description.

{\bf VA platform's requirements on privacy policy.} Application developers are often required to provide a privacy policy and notify users of their apps' privacy practices. VA platforms have different requirements regarding the privacy policies of voice-apps. Google Assistant requires every action to have a privacy policy provided on submission. Amazon Alexa requires skills that collect personal information only to mandatorily have a privacy policy. Both Amazon and Google prevent the submission of a voice-app for certification if their respective requirements are not met~\cite{PolicyGoogleActions, PolicyTesting}. In addition to the privacy policy URL, both platforms offer an option for developers to provide a URL for the terms of use as well. These URLs, if provided by the developers, are made available along with the voice-app's listing on the store.

{\bf Requirements on specific content in privacy policies.}
Google has a "Privacy Policy Guidance" page~\cite{PrivacyPolicyGuidance} in their documentation for action developers. The guide explains what Google's minimum expectation is for a privacy policy document. According to the guide, the privacy disclosures included in the policy should be comprehensive, accurate and easy to understand for the users. The privacy policy should disclose all the information that an action collects through all the interfaces including the data that is collected automatically. How the collected information is used and who and when the collected information is shared with should be specified. 
Google rejects an action if developers do not provide (or even misspell) the action name, company name, or developer email in the privacy policy. The link should be valid and should also be a public document viewable by everyone. Amazon Alexa doesn't provide a guideline for the privacy policy content in their Alexa documentation.



{\bf Voice-app enablement.} VA users enable official (\ie, developed by VA platforms) or third-party voice-apps to expand the functionality of their devices. Voice-apps can be enabled by saying a simple command through voice or by adding it from the smartphone app. A voice-app for which the developer has requested permission to access user's data sends a permission request to the user's VA companion app on smartphone during enablement. The other voice-apps are  directly enabled. A privacy policy can be accessed either over the VA companion app or through the web. On the contrary, it is not accessible through the VA devices over voice. VA platforms do not require end users to accept a privacy policy or the terms of use of a voice-app before enabling it on their devices. It is left for the users to decide whether to go through the privacy policy of the voice-app they use or not.

\begin{figure*}[!h]
	\begin{center}
		\includegraphics[width=0.75\textwidth]{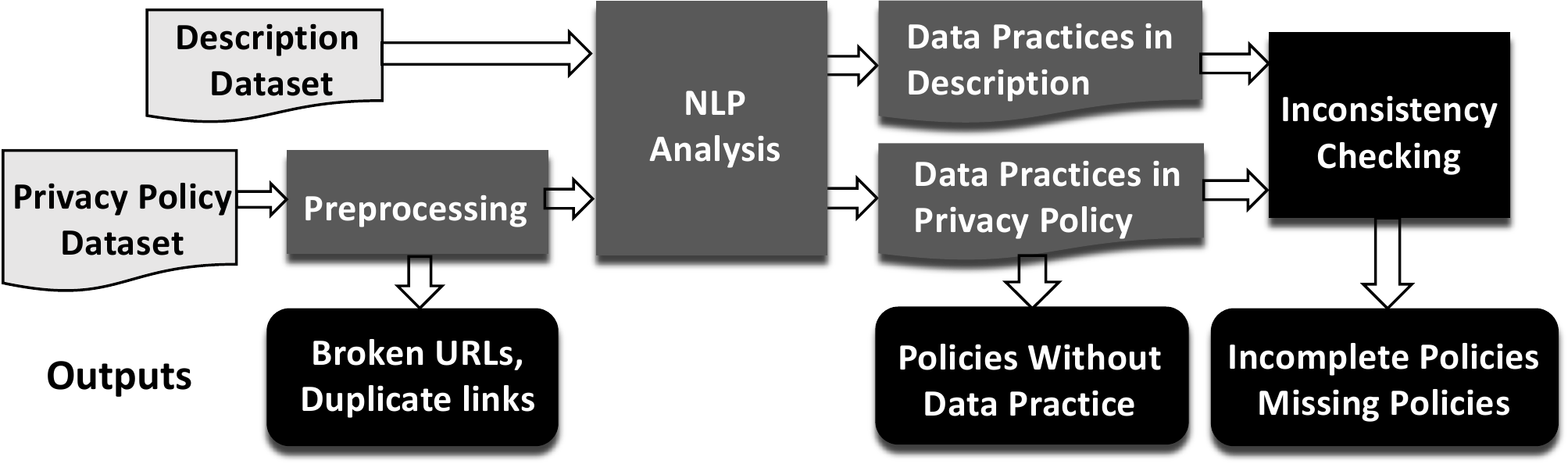}
	\end{center} \vspace{-8pt}
	\caption{Processing pipeline of our privacy policy analysis.}\label{methodologyoverview}
\end{figure*}

\subsection{Challenges on privacy policy analysis} 

Existing privacy policy analysis on smartphone platforms~\cite{Slavin:ICSE:2016, Yu:2016:DSN, zimmeckEtAlCompliance2017, Wang:2018:ICSE, Zimmeck:PETS:2019, SEC:Benjamin:2019} typically conduct static code analysis to analyze potential inconsistencies between an app's privacy policy and its runtime behavior. 
Unlike smartphone app platforms, the source code of voice-apps in Amazon Alexa and Google Assistant platforms are not publicly available. A voice-app is hosted in a server selected by its developer and only the developer has access to it. As far as we know, the source code is not available even to the VA platform's certification teams. This limits the extent of our privacy analysis since we neither have a ground truth to validate our findings of inconsistent privacy policies nor the actual code to find more inconsistencies with the privacy policies provided. The only useful information that we have about a voice-app is the description that is provided by the developer. Descriptions do not have a minimum character count and developers can add a single line description or a longer description explaining all functionalities and other relevant information. Regardless, due to the unavailability of other options, we use the voice-app descriptions for our analysis to detect problematic privacy policies.
For this reason, our results on the inconsistency checking of privacy policies (in Section~\ref{inconsistency}) are not focused on the exact number of mismatches and errors but on the existence of problems potentially affecting the overall user experience.

%% file: datacollection.tex
\section{Methodology}\label{Methodology}

In this section, we first present an overview of our approach, and then detail the major modules including data collection process~(Section~\ref{dataset}), capturing data practices based on NLP analysis~(Section~\ref{datapractices}), and inconsistency checking (Section~\ref{inconsistency}). We seek to understand whether developers provide informative and meaningful privacy policies as required by VA platforms. Fig.~\ref{methodologyoverview} illustrates the processing pipeline of our privacy policy analysis. 
As previously mentioned, each skill/action's listing page on the store contains a description and a privacy policy link (URL). 
We first collect all these webpages, and pre-process them to identify high-level issues such as broken URLs and duplicate URLs. Then, we conduct an NLP based analysis to capture data practices provided in privacy policies and descriptions. 
We seek to identify three types of problematic privacy policies: i) without any data practice; ii) incomplete policies (\eg, a skill's privacy policy lacks data collection information but it has been mentioned in the skill's description); and iii) missing policies (\eg, a skill without a privacy policy but requires one due to its data collection practices). 



\subsection{Data collection}\label{dataset}






We built a crawler to collect a voice-app's id, name, developer information, description and privacy policy link from Amazon Alexa's skills store and Google Assistant's actions store. 
There were several challenges for crawling introduction pages of voice-apps.
First, for the skills store, 23 categories of skills are listed but these are not mutually exclusive. The category "communication" is a subcategory in the "social" category and the category "home services" is a subcategory in the "lifestyle" category. Some skills are classified and listed in multiple categories. We need to remove duplicates during the data collection.
Second, Alexa's skills store only provides up to 400 pages per category, and each page contains 16 skills. Though Amazon Alexa claimed there are over 100,000
skills on its skills store, we were able to crawl only 64,720 unique skills.
Third, Google actions store lists actions in pages that dynamically load more actions when users reach the end of the page. We were unable to use the crawler to automatically get information about all the actions.
As a result, we crawled 2,201 actions belonging to 18 categories from the Google actions store. The total numbers of skills and actions by category we collected are listed in Table~\ref{table:skilldataset} and Table~\ref{table:actiondataset} in Appendix.



Another challenge was to obtain the privacy policy content. Given the privacy policy links, we observed that there are five types of policy pages: i) normal html pages; ii) pdf pages; iii) Google doc and Google drive documents; iv) txt files; and v) other types of files (\eg, doc, docx or rtf). For normal html pages, we used the webdriver~\cite{Selenium} tool to collect the webpage content when they are opened. For the other types of pages, we downloaded these files and then extracted the content from them. Finally, we converted all the privacy policies in different formats to the txt format.

{\bf Privacy Policy Dataset.}
We collected 64,720 unique skills under 21 categories from Alexa's skills store and 17,952 of these skills provide privacy policy links.
Among the 2,201 Google actions we collected, 1,967 have privacy policy links. 

\begin{figure} [!h]
	\begin{center}
		\includegraphics[width=0.45\textwidth]{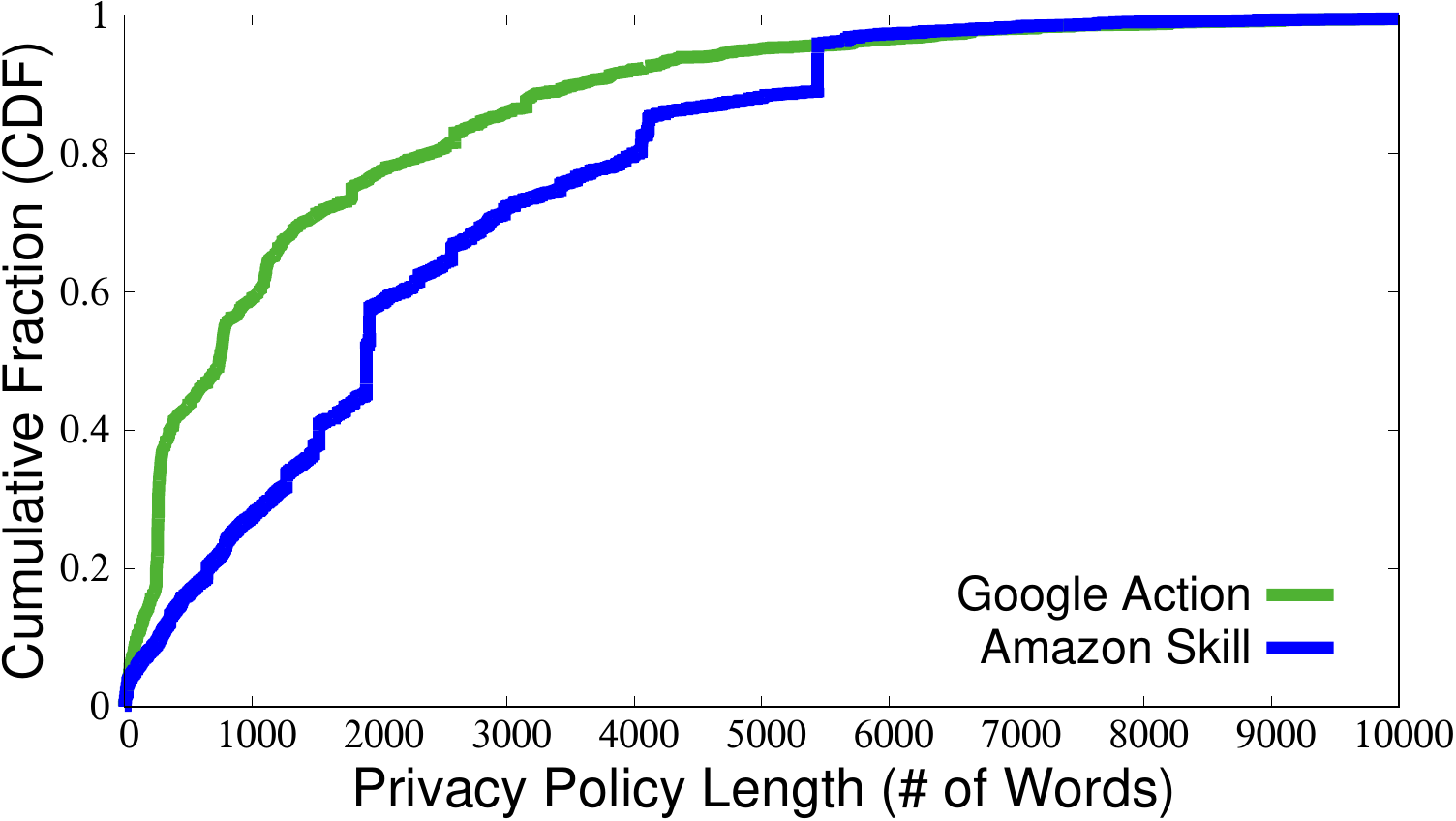}
	\end{center} 
	\caption{Length of a privacy policy.}\vspace{-10pt}	
	\label{cdf:wordlength}
\end{figure}

For each skill/action with a valid policy link, we calculated the number of words in the document. Fig.~\ref{cdf:wordlength} shows the cumulative distribution function of the privacy policy length. The average length is 2,336 words for Alexa skills and 1,479 words for Google actions, respectively. We also observed many very short privacy policies which are not informative. An example is the Google action "Mister Baavlo" which says "{\it We do not store any of your data}" but does not mention what data it collects. Examples of short privacy policies are listed in Table~\ref{table:shortpolicies}.


\begin{table}[h]
	\vspace{-6pt}
	\centering
	\resizebox{8.5cm}{!}{
		{
			\begin{tabular}{ | >{\centering\arraybackslash}m{3.3cm} |>{\centering\arraybackslash}m{2.0cm} | >{\centering\arraybackslash}m{7cm} |}
				\bottomrule[1.5pt]
				\rowcolor[gray]{0.9}Voice-app name & Skill/Action & Privacy policy\\
				\hline
				Passive income tips & Skill & "This is just a sample privacy policy link, You can use this url, If you do not have it."\\
				\hline
				Activity Book &  Skill & "This skill does not collect or save any personal information."\\ 			
				\hline
				BestDateCalendar  &  Skill  & It directs to Google home page\\ 	
				\hline
			    Story Time & Skill (Kids) & " No information is collected during the use of Story Time" \\
				\hline
				KidsBrushYourTeethSong &  Skill (Kids)  & "Privacy Policy" (no content)\\ 
				\hline

				Mister baavlo   & Action &  "We do not {store} any of your data" \\
				\hline 
				Sanskrit names for yoga poses   & Action &  "Google Docs: You need permission to access this published document." \\
				\bottomrule[1.5pt]
			\end{tabular} 
		}
	}
	\vspace{-0pt}
	\caption{Examples of short privacy policies.}
	\label{table:shortpolicies}\vspace{-20pt}
\end{table}

{\bf Description Dataset.} Description of a voice-app is intended to introduce the voice-app to end users with information regarding its functionality and other relevant information. It may also contain data practices (\eg, the data required to be collected to achieve a functionality) of the voice-app. We collected voice-app descriptions and used them as baselines to detect potentially inconsistent privacy policies. In our dataset, all skills/actions come with descriptions. 



\subsection{Capturing data practices}\label{datapractices}

In order to automatically capture data practices in privacy policies and descriptions of voice-apps, we develop a keyword-based approach using NLP. However, we want to emphasize that we do not claim to resolve challenges for comprehensively extracting data practices (\ie, data collection, sharing, and storing) from natural language policies. Instead, we mainly focus on obtaining empirical evidences of problematic privacy policies using a simple and accurate (\ie, in terms of the true positive) approach. We discuss the limitation of our approach in Section~\ref{Discussion}.

{\bf Verb set related to data practices.} 
Researchers in~\cite{BHR13,Yu:2016:DSN} have summarized four types of verbs commonly used in
privacy policies: \texttt{Collect}, \texttt{Use}, \texttt{Retain} and \texttt{Disclose}. Each type contains semantically similar verbs in terms of the functionality. \texttt{Collect} means an app would collect, gather, or acquire data from user; \texttt{Use} indicates an app would use or process data; \texttt{Retain} means storing or remembering user data; and \texttt{Disclose} means an app would share or transfer data to another party. 
\begin{table}[h!]
		\vspace{-6pt}
	\centering
	\resizebox{8.5cm}{!}{
		\begin{tabular}{  >{\centering\arraybackslash}m{1.0cm} >{\centering\arraybackslash}m{8.9cm} }\toprule[1.5pt]
		{\bf Verb Set} 	& Access, Ask, Assign, Collect, Create, Enter, Gather, Import, Obtain, Observe, Organize, Provide, Receive, Request, Share, Use, Include, Integrate, Monitor, Process, See, Utilize, Retain, Cache, Delete, Erase, Keep, Remove, Store, Transfer, Communicate, Disclose, Reveal, Sell, Send, Update, View, Need, Require, Save  \\
			
			\bottomrule[1.0pt]

	{\bf 	Noun Set} & Address, Name, Email, Phone, Birthday, Age, Gender, Location, Data, Contact, Phonebook, SMS, Call, Profession, Income, Information\\
			\bottomrule[1.5pt]
		\end{tabular} 
	}
	\vspace{-0pt}
	\caption{Keyword dictionary related to data practices.}\label{table:Verb}
	\vspace{-15pt}
\end{table}


{\bf Noun set related to data practices.} 
From Amazon's skill permission list~\cite{ConfigurePermissions} and Amazon Developer Services Agreement~\cite{AmazonDeveloperAgreement}, we manually collected a dictionary of 16 nouns related to data practices. Table~\ref{table:Verb} lists a dictionary with 40 verbs and 16 nouns that we used in our privacy policy analysis.

{\bf Phrases extraction.} We first parsed a privacy policy into sentences. We used the SpaCy library~\cite{spacy} to analyze each sentence, and obtained the attribute for each word. SpaCy can effectively find the straight correlation between a noun and a verb and ignore other words in a sentence. We identified three types of basic phrases:
\begin{itemize}
\item noun (subject) + verb, \eg, "Alexa (will) tell" or "email (is) required"
\item verb + noun (object), \eg, "send (a) message"
\item verb + noun (object) + noun + noun, \eg, "tell (you) (the) name (of) meeting (on) (your) calendar"
\end{itemize}

Next, we combined two basic phrases to generate a longer phrase if they share the same verb. The combined phrase would follow patterns: "subject+verb+object" or "subject+is+passive verb". For example, for a sentence "Alexa skill will quickly tell you the name and time of the next meeting on your Outlook calendar", we obtained the phrase "Alex tell name, meeting, calendar". 

{\bf Identifying data practices.} Given all phrases extracted from the privacy policy and description, we used the verb and noun sets in Table~\ref{table:Verb} to identify data practice phrases. Specifically, we compared the similarity of verb and noun in each phrase with the verb and noun sets respectively. If the similarity is higher than a threshold, we consider the phrase as a data practice phrase. To measure the semantic similarity of two phrases, we used the similarity measurement based on word2vec provided by SpaCy, which is determined by comparing two word vectors. We set the similarity threshold to 0.8 in our analysis, so as to achieve a high true positive rate (but no guarantee of the false negative rate). For example, our tool identified privacy policies of 1,117 skills in Amazon Alexa and 95 actions in Google Assistant having zero data practice (details are in Sec.~\ref{ZeroDP}). Our manual analysis confirmed that these privacy policies do contain no data practice.

\subsection{Inconsistency checking}\label{inconsistency}

With description phrases and privacy policy phrases for each voice-app, we checked any potential inconsistency between them. First, if the data practice phrases in a description are not semantically similar to any data practice phrase in the corresponding privacy policy, we consider this privacy policy to be incomplete. 
For example, the description of skill "Thought Leaders" mentions "Permission required: Customer's Full name, Customer's Email Address, Customer's Phone number", but none of them are mentioned in its privacy policy. We consider it an incomplete privacy policy.

Second, since Amazon Alexa only requires skills that collect personal information to provide a privacy policy, we detected whether a privacy policy of an Alexa skill is missing although it is required. If the description mentions that a skill collects some data but the skill has no privacy policy, we consider that the skill lacks a privacy policy.
For example, a skill "Heritage Flag Color" mentions "The device location is required" in its description. But the developer doesn't provide a privacy policy. Note that it only reflects an inconsistency between the privacy policy and description since there is a lack of the ground truth to validate whether the skill really collects the location information or not.

%% file: policyanalysis.tex
\section{Major Findings}\label{Findings}
In this section, we discuss major findings from our analysis of the privacy policies available in the stores of both Amazon Alexa and Google Assistant. 
We first present high-level issues such as broken and incorrect privacy policy URLs, duplicate privacy policy links, and issues in Google and Amazon's official voice-apps. Then, we conduct a content analysis of privacy policies, and discuss the issues such as zero data practice and inconsistency in privacy policies.    
In addition, we discuss usability issues of privacy policies for voice-apps. We back our findings with representative examples that we found from the app stores during our analysis. 

\subsection{High-level issues}\label{sec:Finding1}

\begin{table}[!htb]
	\label{eva:overhead}
	\begin{center}
		\resizebox{8.0cm}{!}{
			\begin{tabular}{ >{\centering\arraybackslash}m{3.7cm} >{\centering\arraybackslash}m{1.2cm}  >{\centering\arraybackslash}m{1.2cm} >{\centering\arraybackslash}m{1.2cm} >{\centering\arraybackslash}m{1.2cm}   }
				\cline{1-5}
				& \multicolumn{2}{c}{Amazon skills} & \multicolumn{2}{c}{Google actions}   \\
				\cline{2-5}
				& Total \# & Percentage & Total \# & Percentage \\
				\hline
				    Without privacy policy & 46,768 & 72\% & 234 & 11\%   \\
			    \hline
			        Valid privacy policy URL & 16,197 & 25\% & 1,887 & 85\%  \\
				\hline
				    Broken privacy policy URL & 1,755 & 3\%  & 80 & 4\%\\
				\hline
			\end{tabular}
		}
	\end{center}
	\caption{Statistics of privacy policies on two VA platforms.}
	\label{AlexaDataOverview}
\end{table}

\subsubsection{Not all voice-apps have a privacy policy URL}\label{sec:Finding1:missing}
Both Google and Amazon have taken different approaches when it comes to the requirement of a privacy policy for each voice-app available to users. While Google has made it mandatory for developers to provide a privacy policy along with each action, Amazon is more lenient and makes it a requirement only for skills that declare that they collect personal information through the skill. On analyzing the stores, we have noticed irregularities concerning this, as illustrated in Table~\ref{AlexaDataOverview}. Out of the 2,201 actions we collected from the Google action directory, only 1,967 have privacy policies provided which means that 11\% of the actions do not have a privacy policy provided. While it is not possible to submit an action for certification without including a privacy policy URL, it is puzzling how these actions are available in the store without providing one. Out of these 234 actions that do not have privacy policies, 101 actions were developed by Google while the other 133 actions were developed by 41 other developers.

In the case of Alexa skills, as shown in Table~\ref{AlexaDataOverview}, only 17,952 (28\%)  skills have a privacy policy out of the 64,720 skills we collected (\ie, 46,768 skills without a privacy policy). It is partially because of the lenient skill certification in Alexa. After conducting further experiments on the skill certification, we have understood that even if a skill collects personal information, the developer can choose to not declare it during the certification stage and bypass the privacy policy requirement. This is achieved by collecting personal information through the conversational interface (\eg, asking users' names). Even though this data collection is prohibited, the certification system of Amazon Alexa doesn't reject such skills.
As a result, developers may choose to not provide a privacy policy. Amazon only requires skills that collect personal data to provide a privacy policy, and thus not all these 46,768 skills require a privacy policy. In Sec.~\ref{missingpolicy}, we identify skills that potentially lack a required privacy policy.

\subsubsection{Broken links and incorrect URLs} 
\begin{figure} [!h]
	\vspace{-10pt}
	\begin{center}
		\includegraphics[width=0.35\textwidth]{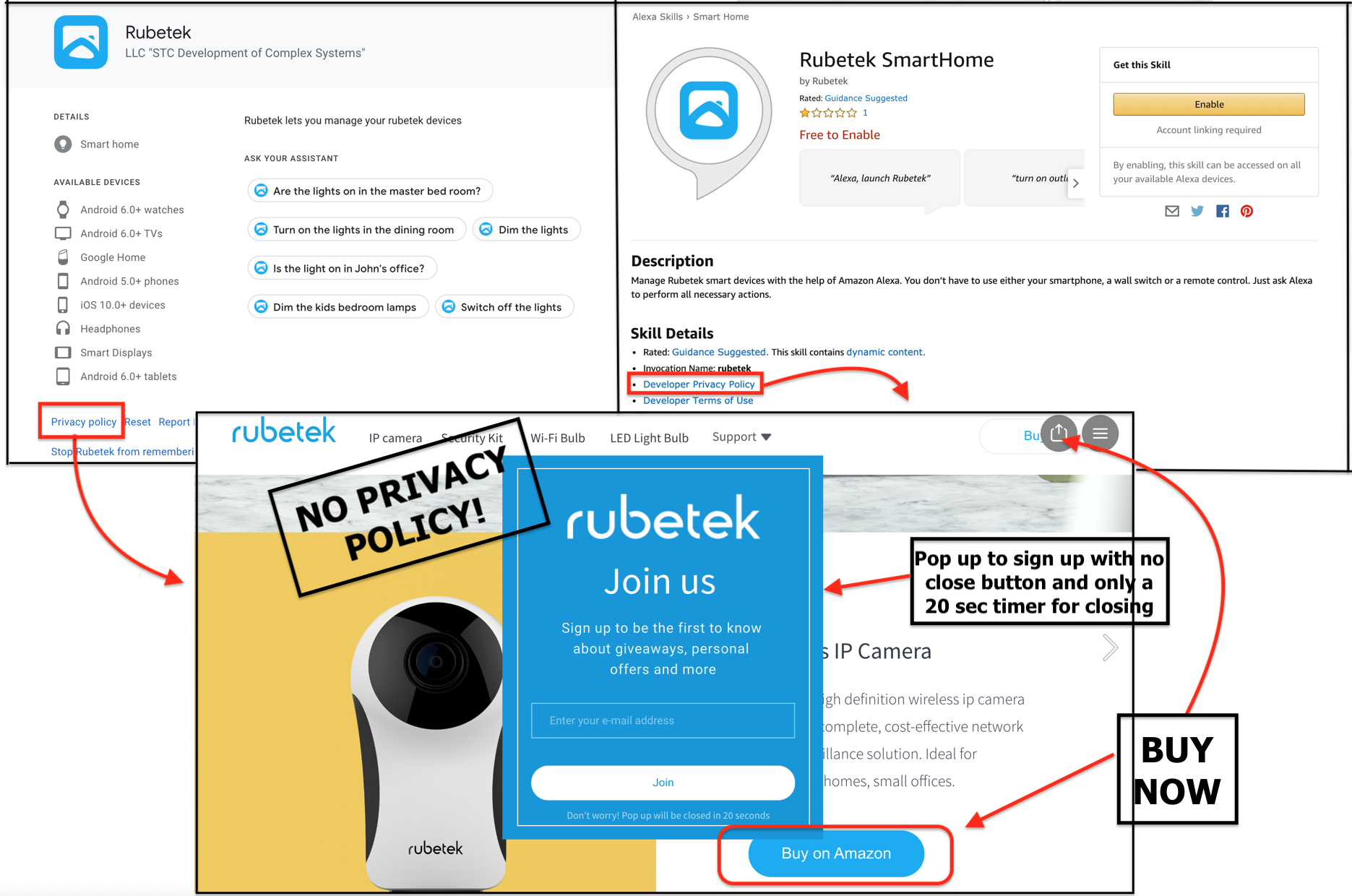}
	\end{center} 
	\vspace{-10pt}
	\caption{Landing page of the privacy policy URL provided with the Google action and Alexa skill developed by Rubetek.}
	\label{promotion}
    \end{figure}

For those actions and skills that have provided a privacy policy URL, not every URL leads to the page containing a privacy policy. Through our experiments, we found 80 Google actions and 1,755 Alexa skills that have provided broken privacy policy URLs, as shown in Table~\ref{AlexaDataOverview}. 
There are also URLs which lead to other developer's privacy policies. An example for this is the skill "NORAD Tracks Santa" by NORAD which provides a privacy policy URL that links to Amazon's privacy policy page instead of a privacy policy written by the developer. The privacy policy URL of "Rubetek SmartHome" which is both an Alexa skill and a Google action leads to the company's homepage which promotes its products, as shown in Fig.~\ref{promotion}, rather than linking to the privacy policy page. 
Sec.~\ref{sec:Finding2} presents our  content analysis of privacy policies, which provides more details about the voice-apps with incorrect privacy policy URLs.


\subsubsection{Duplicate URLs}
\begin{figure} [!h]
	\begin{center}\vspace{-12pt}
		\includegraphics[width=0.45\textwidth]{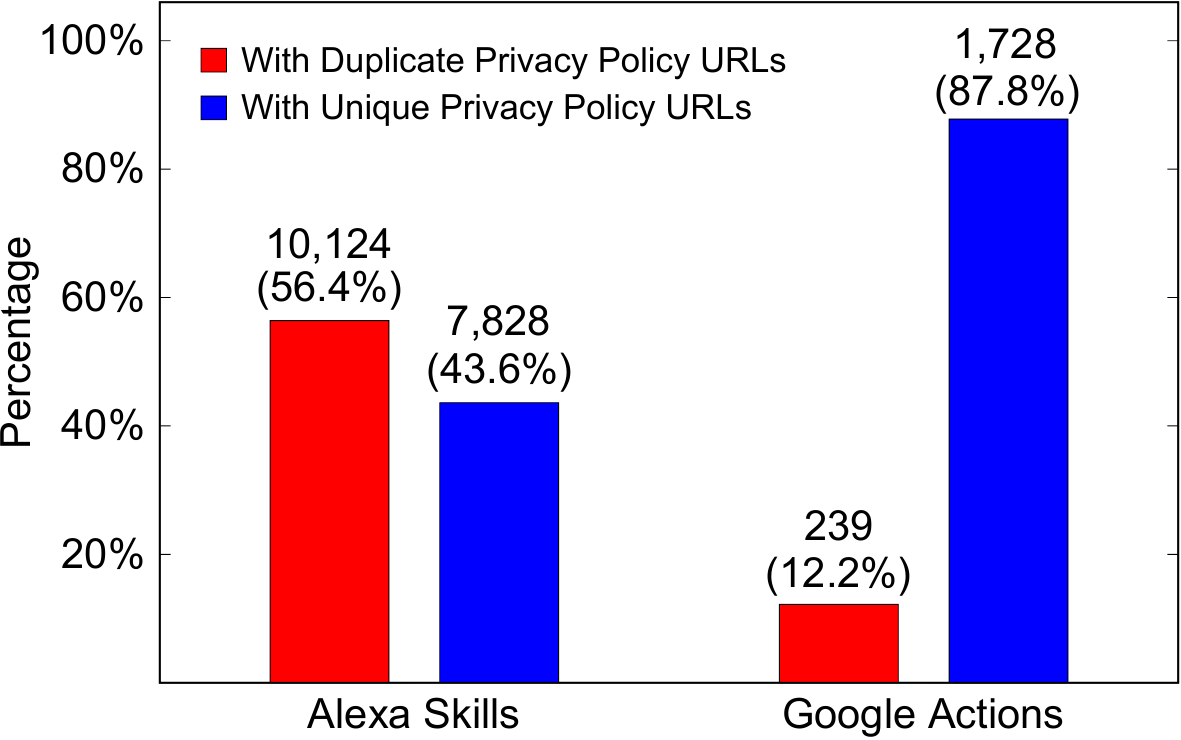}
	\end{center} 
	\vspace{-10pt}
	\caption{Duplicate privacy policy URLs in two platforms.}
	\label{DuplicateLink}
\end{figure}

We found a substantial portion of privacy policies share same URLs. In particular, Amazon Alexa has more than 56\% of skills with duplicate privacy policy URLs.
Fig.~\ref{DuplicateLink} shows the prevalence of duplicate privacy policy URLs in both platforms. Out of the 17,952 Amazon skills with privacy policies, 7,828 skills have a unique privacy policy URL. The other 10,124 skills (56.4\%) share 1,206 different privacy policy URLs. Out of these, 1,783 skills (9.9\%) have provided the same link (\url{https://getstoryline.com/public/privacy.html}) as their privacy policy URLs. Note that these 1,783 skills are not from the same developer which indicates that the privacy policy is irrelevant to these skills.
Table~\ref{table:duplicate} lists the most common privacy policy URLs in Alexa and Google platforms. The issue of duplicate URLs is more serious on Amazon Alexa platform. The top three duplicate URLs are shared by 3,205 skills, constituting 17.8\% of the total skills that have a privacy policy. 
As shown in Fig.~\ref{DuplicateLink}, the Google Assistant platform has 12.2\% of actions with duplicate privacy policy URLs. 1,728 out of 1,967 actions have a unique privacy policy. The other 239 actions share 64 different privacy policy URLs. 

\begin{table}[!htb]
\vspace{-5pt}
\centering
	\resizebox{8.0cm}{!}{
		{
			\begin{tabular}{  >{\centering\arraybackslash}m{1cm} >{\centering\arraybackslash}m{5.5cm}  >{\centering\arraybackslash}m{1cm}  >{\centering\arraybackslash}m{1.2cm} }
				\hline
                  \rowcolor[gray]{0.9}   Platform &Duplicate privacy policy URLs  &  Total \# &  Percentage \\
				\hline
			\multirow{3}{*}{Amazon}& 
    https://getstoryline.com/public/privacy.html
 & 1,783  &  9.9\% \\
                \cline{2-4}
				
				&https://corp.patch.com/privacy & 1,012 & 5.6\%  \\
			 \cline{2-4}
				&https://cir.st/privacy-policy & 410 & 2.3\% \\
				\hline
				\multirow{3}{*}{Google}&
				https://xappmedia.com/privacy-policy
 & 32  &  1.6\% \\
				\cline{2-4}
				&https://voxion.us/google-privacy-policy & 24 & 1.2\%  \\
			\cline{2-4}
			&	https://www.spokenlayer.com/privacy & 20 & 1.0\% \\
				\hline
				
			\end{tabular}
		}
	}
	\vspace{2pt}
	\caption{{The most common duplicate privacy policy URLs.}}

	\label{table:duplicate}
\end{table}	

To understand why there exists such a large number of voice-apps with duplicate privacy policy URLs especially on Amazon Alexa platform, we further examined the developer information of these voice-apps. Our intuition is that developers who published multiple voice-apps may use the same privacy policy URLs. We found that for the developers who developed more than one skill, 77\% of their skills use duplicate privacy policy URLs. Table~\ref{table:developer} lists the top 5 developers who published the most skills with a privacy policy on Amazon Alexa platform. As illustrated in the table, 2,064 out of 2,069 skills (99.8\%) use duplicate privacy policy URLs. Obviously, the content of these privacy policy URLs are not skill-specific, and users may skip reading the privacy policy although it is provided. 
A serious problem happens if such a privacy policy link is broken, which results in hundreds of skills being affected. For example, we found a broken link "\url{https://www.freshdigitalgroup.com/privacy-policy-for-bots}" (shown in Table~\ref{table:developer}). There are 217 skills using this link, and thus all their privacy policies become inaccessible.
As to the Google actions, we also observed the similar issue. Although Google requires that a privacy policy must include one of the following: action name, company name or developer email, there are developers using a general privacy policy with the company name or email for all their actions. For the developers who published more than one action, 27.5\% of actions have duplicate privacy policy URLs. For the top 10 developers who published the most actions, 86\% of their actions use a duplicate privacy policy link.
 
\begin{table}[!htb]
            	\centering
	\resizebox{8.5cm}{!}{
		{
			\begin{tabular}{ | >{\centering\arraybackslash}m{2.5cm} |>{\centering\arraybackslash}m{1.8cm}  |>{\centering\arraybackslash}m{2.0cm}  |>{\centering\arraybackslash}m{6cm}|}
				\hline
                \rowcolor[gray]{0.9} Developer & \# of skills developed  & Skills with duplicate URLs & Top duplicate URLs used by the developer \\
				\hline
				Patch.com & 1,012 & 1,012 & \url{http://corp.patch.com/privacy}   \\
			    \hline
				Radio.co & 295 & 292 & \url{http://www.lottostrategies.com/script/showpage/1001029/b/privacy_policy.html} \\
				\hline
				Tinbu LLC & 264 & 263 & \url{http://spokenlayer.com/privacy} \\
		        \hline
			    FreshDigitalGroup &	259 & 258 & \url{https://www.freshdigitalgroup.com/privacy-policy-for-bots}  \\
				\hline
			    Witlingo &	239 & 239 & \url{http://www.witlingo.com/privacy-policy} \\
				\hline				
			\end{tabular}
		}
	}
	\vspace{4pt}
	\caption{{Top 5 developers that published the most skills with a privacy policy on Amazon Alexa platform.}}
	\label{table:developer}
\end{table}




\subsubsection{There are Google and Amazon's official voice-apps violating their own requirements.}
\begin{figure} [!h]
\vspace{-10pt}
	\begin{center}
		\includegraphics[width=0.4\textwidth]{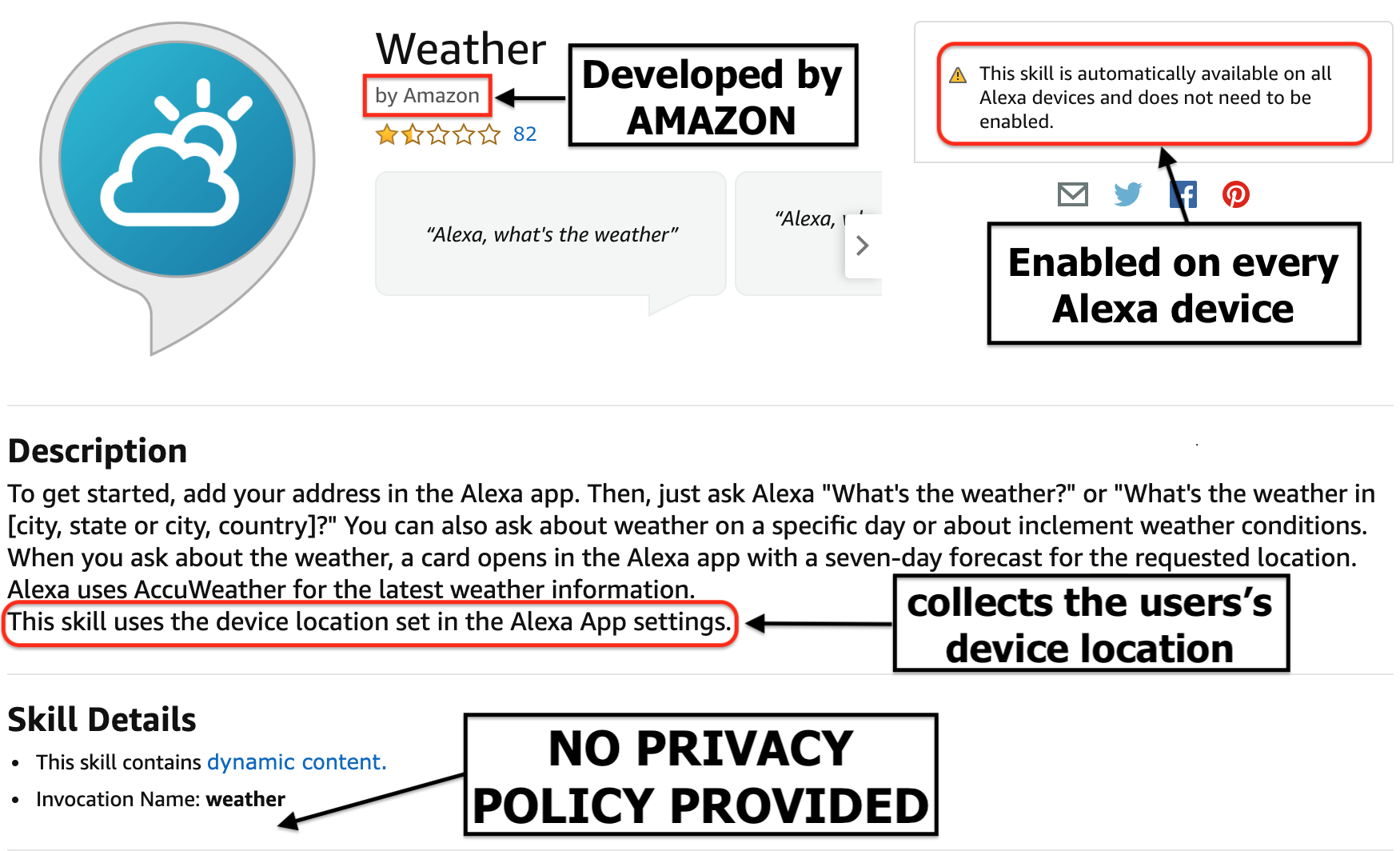}
	\end{center} 
	\vspace{-10pt}
	\caption{An official skill lacks a privacy policy. Even though it collects the user's location according to the description, no privacy policy is provided.}
	\label{weather}
\end{figure}

We found two official "Weather" skills on 
Amazon Alexa's skills store, and one of them asks for user's location according to the description but it doesn't provide a privacy policy. Fig.~\ref{weather} shows the "Weather" skill developed by Amazon. This skill may be automatically enabled and available on all Alexa devices since it is a built-in skill. This example demonstrates that Amazon Alexa violates its own requirement by publishing voice-apps capable of collecting personal information without providing a privacy policy.

We collected 98 Amazon Alexa official skills (\ie, developed by Amazon, Amazon Alexa Devs, and Amazon Education Consumer Team), out of which 59 skills come with privacy policy URLs (but all are duplicate URLs). Among these privacy policy links, 30 links point to the general Amazon privacy notice and 6 links are the AWS (Amazon Web Services) privacy notice, Amazon payment privacy or Alexa term of use. Surprisingly, 23 privacy policy links are totally unrelated to privacy notice, in which 17 links are Amazon homepage and 6 links are pages about insurance.
In Google's actions store, we found 110 official actions developed by Google, in which 101 actions don't provide a privacy policy. For the 9 actions with a privacy policy link, they point to two different Google Privacy Policy pages and both are general privacy policies. Google requires that every action should have an app-specific privacy policy provided by developers on submission. However, our analysis reveals that this requirement has not been enforced in a proper manner.



\subsection{Content analysis of privacy policies}\label{sec:Finding2}

\subsubsection{Irrelevance to specific voice-app}\label{sec:Finding2:general}
It is important to cover all aspects of a service's data practices in the privacy policy. The contradiction is providing these data practices for a service that is not capable of doing any of the data collections mentioned in the privacy policy. This is especially evident in the Alexa skill store where most skills have a privacy policy that is common across all services that the developers provide. These policies do not clearly define what data practices the skill is capable of. Some of these privacy policies do not even mention the Alexa skill or Google action as a service and state that it is the privacy policy of a specific service such as the website. 
We analyzed whether a voice-app mentions the app name in its privacy policy. There are only 3,233 skills out of 17,952 skills (around 18\%) mentioning skills' names in their privacy policies. For Google actions, 1,038 out of 1,967 actions (around 53\%) mention action names in the privacy policies. 

\begin{figure} [!h]
	\begin{center}
		\includegraphics[width=0.45\textwidth]{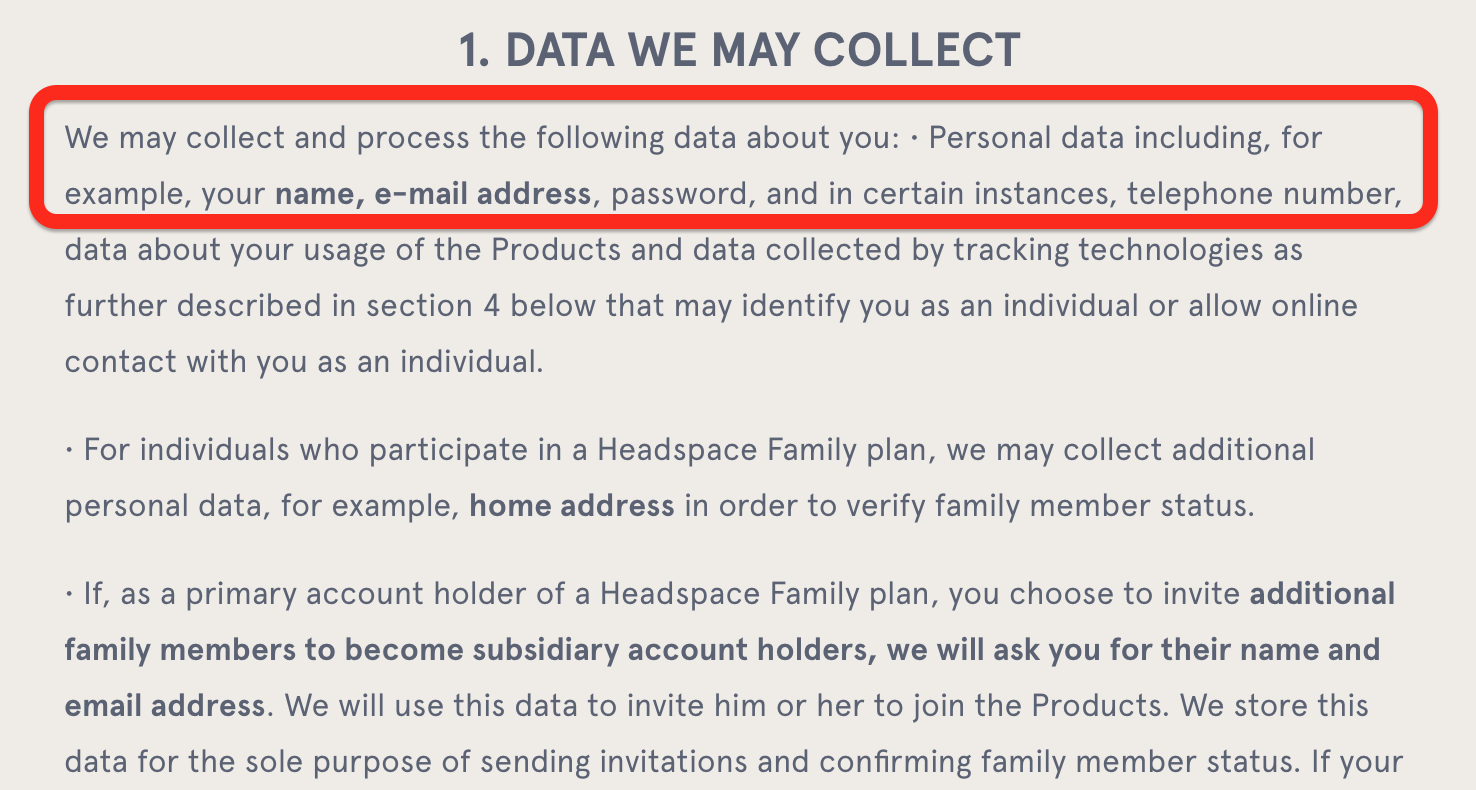}
	\end{center} 
	\vspace{-10pt}
	\caption{Privacy policy URL provided with a kids skill "Headspace Bedtime Story" disclosing the collection of personal data which is prohibited according to Amazon Alexa's privacy requirements~\cite{PrivacyRequirements}.}
	\label{irrelevant}
\end{figure}
    
There were also privacy policies provided for kids skills which mention that the service is not intended to be used by children and also that the service can collect some form of personal information, which is not allowed for skills in kids category according to Amazon Alexa's privacy requirements~\cite{PrivacyRequirements}. Fig.~\ref{irrelevant} shows an example where the privacy policy URL provided with a kids skill disclosing the collection of personal data. In addition, we found 137 skills in the Amazon Alexa's kids category whose privacy policies mention data collection is involved. But they just provide a general privacy policy. All these skills potentially violate Amazon Alexa’s privacy requirements on kids skills.

\subsubsection{Zero data practice}\label{ZeroDP}
\begin{figure} [!h]
\vspace{-6pt}
	\begin{center}
		\includegraphics[width=0.45\textwidth]{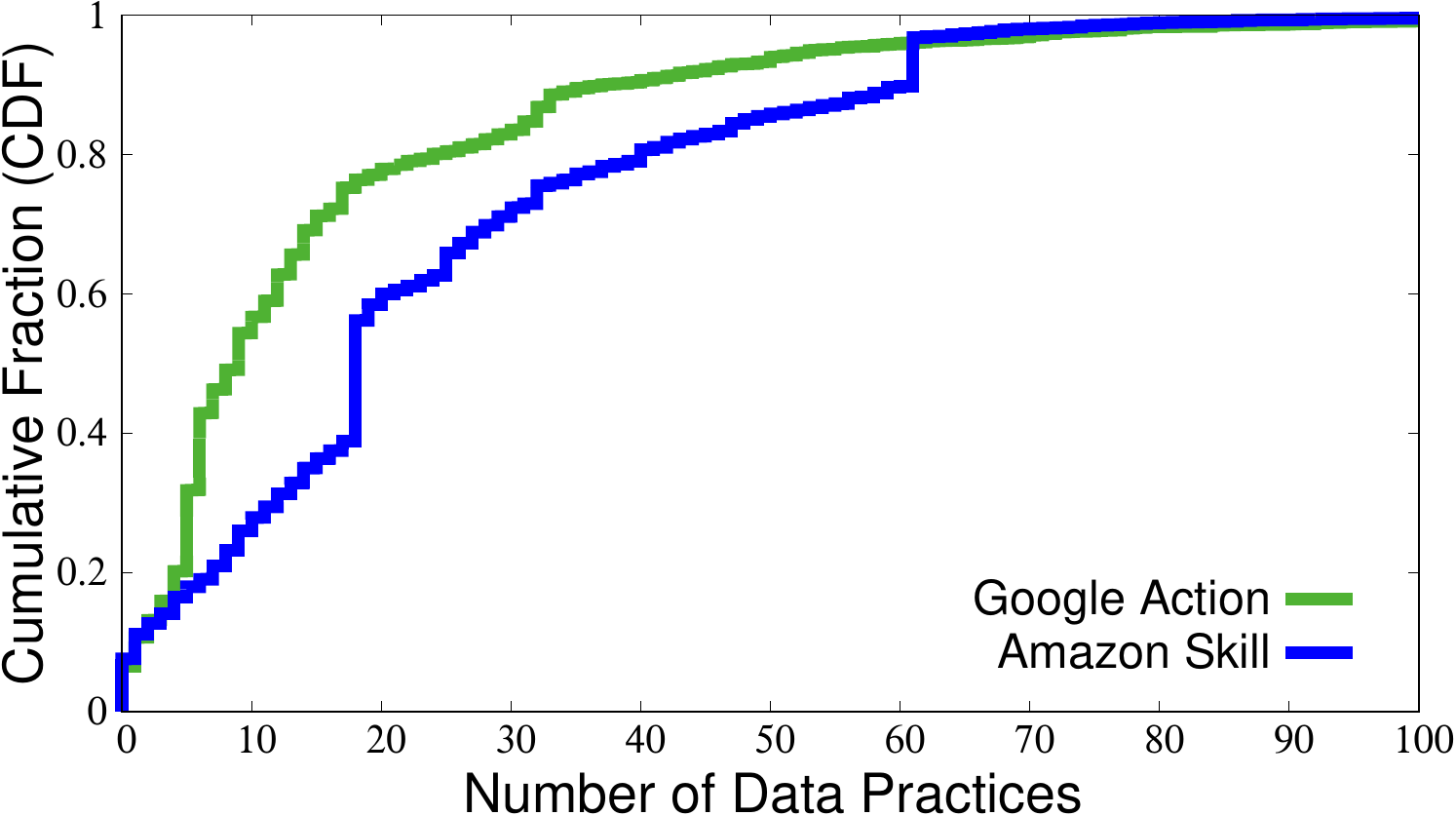}
	\end{center} 
	\vspace{-10pt}
	\caption{Number of data practices in a privacy policy.}
	\label{cdf:datapractices}
\end{figure}

We applied our method described in Sec.~\ref{datapractices} to capture data practices in each privacy policy.
Fig.~\ref{cdf:datapractices} illustrates the cumulative distribution function of data practices we identified using our privacy policy dataset. For these privacy policies with data practices, the average amount is 24.2 in Amazon Alexa and 16.6 in Google Assistant, respectively. The maximum number of data practices in a privacy policy is 428, which is likely a general privacy policy rather than an app-specific one.

\begin{figure} [t]
	\begin{center}
    \subfigure[Amazon skills with no data practice]{\includegraphics[width=0.225\textwidth]{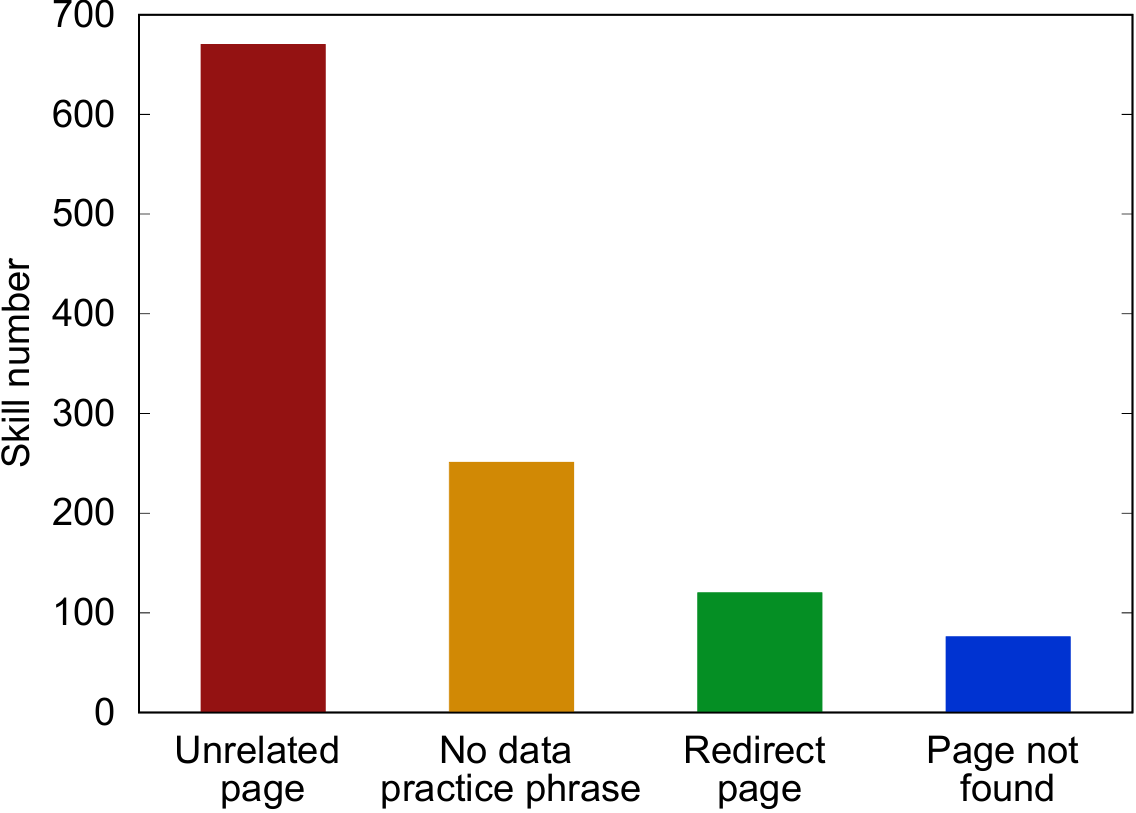}}
    \subfigure[Google actions with no data practice]{\includegraphics[width=0.23\textwidth]{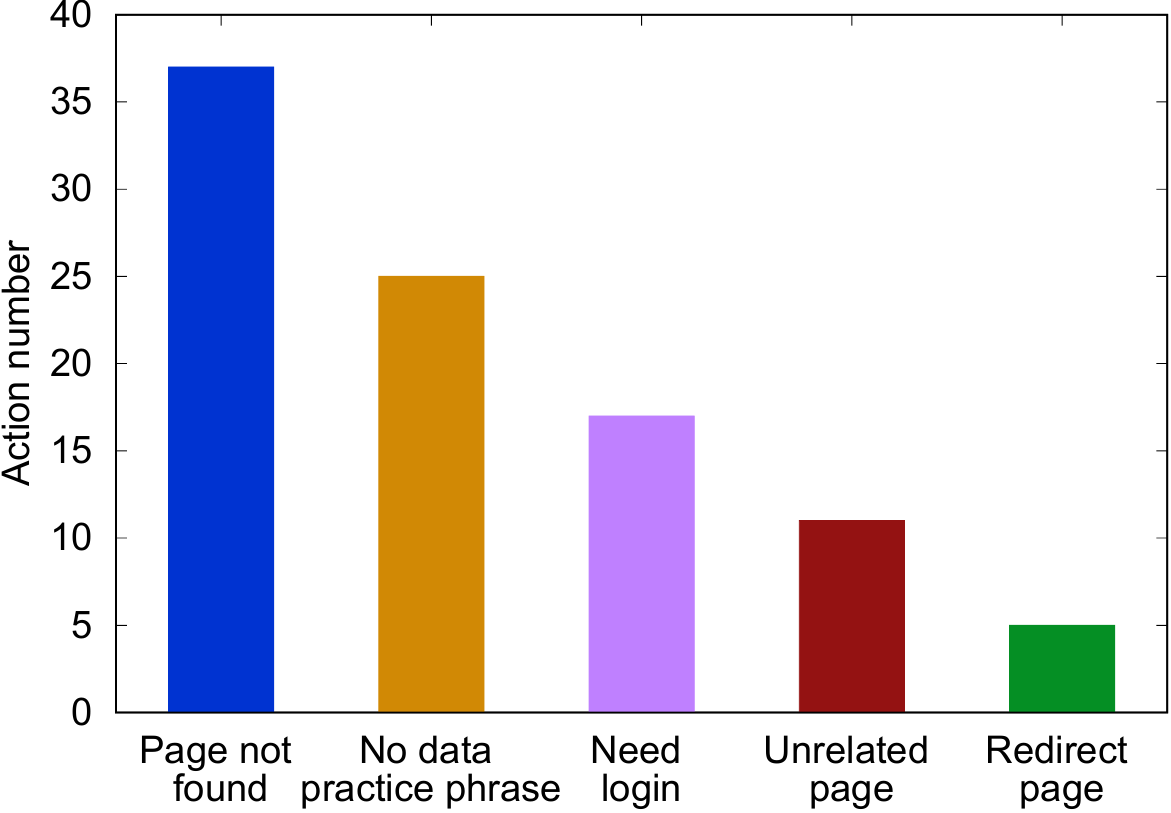}}
	\end{center} 
	\vspace{-12pt}
	\caption{Different issues of privacy policies that have zero data practice in two VA platforms.} 
	\label{Alexadatapracticeissues}
\end{figure}

In particular, 1,117 privacy policies provided with Alexa skills have zero data practices. Fig.~\ref{Alexadatapracticeissues} shows the breakdown of different issues of these privacy policies. 670 URLs lead to totally unrelated pages which have advertisements and shopping options. 251 URLs lead to an actual privacy policy page but has no data practices mentioned. 
120 URLs lead to a page where the actual link to the privacy policy does exist but will be redirected to some other pages. 76 URLs lead to an actual website domain but the link is not found. These too can be considered as broken links.

Our tool identified 95 Google actions having privacy policies with zero data practice, as shown in Fig.~\ref{Alexadatapracticeissues}. 37 URLs lead to a page that is not found. 25 URLs are privacy policies but with no data practice. 11 URLs lead to unrelated links with shopping options and product advertisements. 5 URLs lead to a page containing the link to the actual privacy policy. In addition, 17 actions provide their privacy policy as a Google doc which does not have the correct permissions set which results in users not being able to access it. This was exclusively found within the Google actions while they violate Google's restriction "the link should be a public document viewable by everyone". 

\subsubsection{Inconsistency between the privacy policy and description}\label{Inconsistencychecking}

\begin{figure} [!h]
\vspace{-10pt}
	\begin{center}
		\includegraphics[width=0.45\textwidth]{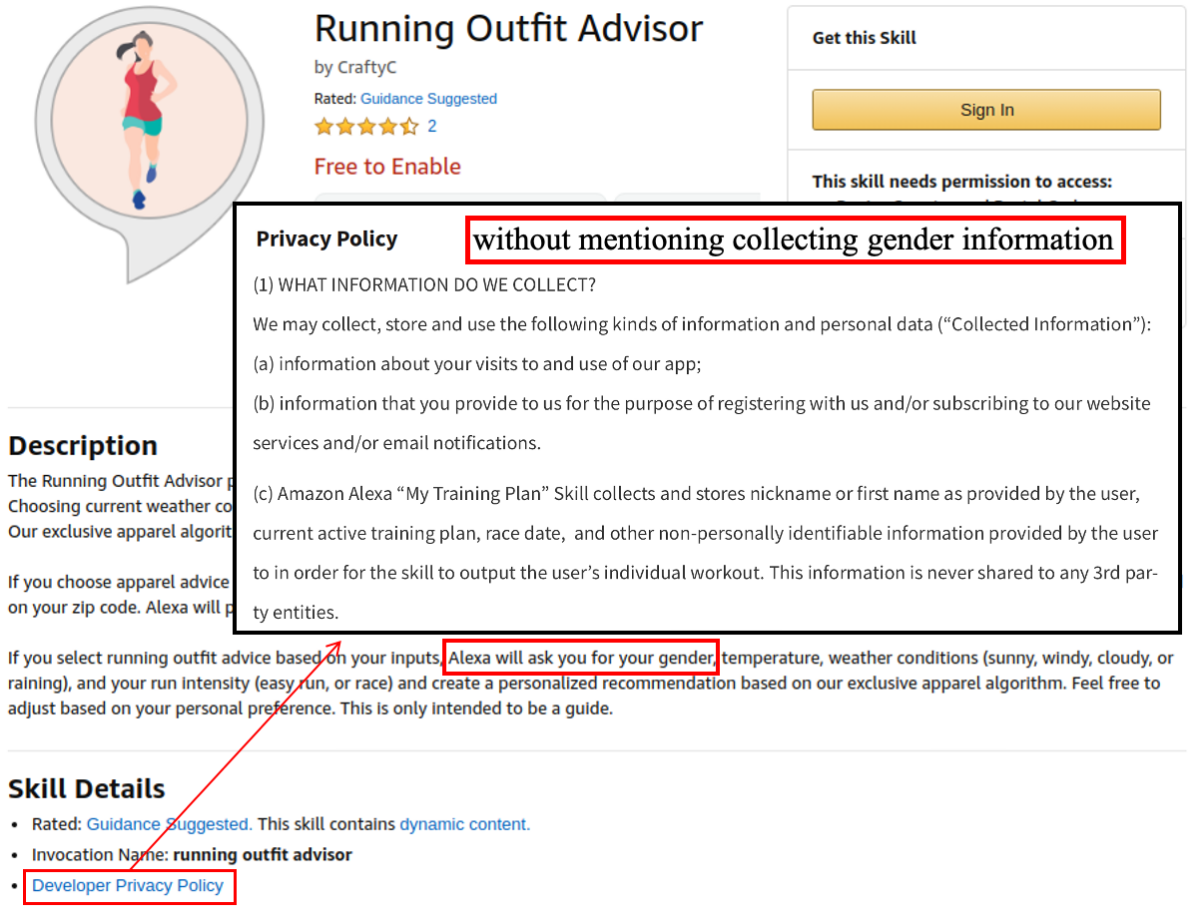}
	\end{center} 
	\vspace{-10pt}
	\caption{"Running Outfit Advisor" skill mentions collecting the gender information in the description, but does not mention this data practice in its privacy policy.}
	\label{noprivacy2}
\end{figure}

Using the method presented in Section 3.3, we identified 50 Alexa skills that have a privacy policy which is inconsistent with the corresponding description. These skills describe the collection of personal data in the description but these data practices are not mentioned in the privacy policy provided. The consequence of such occurrences is that the users are not informed about what happens to their information and who it is shared with. Among the 50 skills, 19 skills ask for address or location; 10 skills request email/ account/ password; name is asked by 7 skills and 4 skills require the birthday; other skills ask for phone number, contact, gender or health data.
Fig.~\ref{noprivacy2} shows an example, where the skill "Running Outfit Advisor" mentions collecting the gender information in the description, but does not mention this data practice in its privacy policy. 
In another case, the description of the skill "Record - Journal - Things to Do Calendar"  describes the collection of personal information like the address of the user. The description has the following line: "{\it Device Address. Your device address will be used to provide responses with events local to your area.}" In the skill's privacy policy, the data practices are not disclosed clearly enough but only says "we will collect personal information by lawful". We treated this kind of privacy policy as inconsistent (incomplete) privacy policy since it fails to give a clear idea about its data practices.
Table~\ref{table:incompletepolicy} in Appendix shows the list of these skills with inconsistency between the privacy policy and description.

\subsubsection{Missing required privacy policies}\label{missingpolicy}
In Sec.~\ref{sec:Finding1:missing}, we have shown 
234 Google actions do not have a privacy policy provided, which violates its own restriction "Google require all actions to post a link to their privacy policy in the directory". Here we focus on Amazon Alexa skills and identify cases with missing required privacy policies.

To collect user's personal data for use within the voice-apps, developers can use the built-in feature of collecting the personal information directly from their Amazon account after taking permissions from the user. This permission is taken from the user when the skill is first enabled. While this is appropriate and respect the users privacy, there is another channel that can be misused for collecting personal information. A developer can develop a skill to ask for the personal information from the user through the conversational interface. Both Amazon and Google prohibit the use of conversational interface to collect personal data. But, in the case of Amazon, this is not strictly enforced in the vetting process. By collecting personal information in this manner, the developer can avoid adding a privacy policy URL to the skill's distribution requirements. This is possible because Amazon requires only skills that publicly declare that they collect personal information to mandatorily have a privacy policy. The developer can easily bypass this requirement by lying about not collecting personal information.
\begin{table}[h]
	\vspace{-5pt}
	\centering
	\resizebox{8.0cm}{!}{
		{
			\begin{tabular}{ | >{\centering\arraybackslash}m{2cm} | >{\centering\arraybackslash}m{1.5cm} | >{\centering\arraybackslash}m{7cm} |}
				\hline
				\rowcolor[gray]{0.9} Data &  \# of Skills &  Skills names \\
				\hline
				Name & 10 & First Name Analysis, Haircut Scheduler, insurance service, LOVE CALCULATOR, Mr. Tongue Twister (Kids), My daily task, Name My Grandkids, Social Network, Uncle Tony (Kids), who's right \\
				\hline
				Location & 6  &  Doctor Locator,  Heritage Flag Color, Lapel Athletics, OC Transpo, Weather, World Time \\
				\hline
				Gender & 1 & Interactive Bed Time Story (Kids) \\
				\hline
				Age & 1 & bright smile  \\
				\hline
				Birthday & 1 & Cake Walk \\
				\hline
				Ip Address & 1 & Network-Assistant \\
				\hline
			\end{tabular}
		}
}
	\vspace{4pt}
	\caption{{Skills with no privacy policies despite mentioning the collection of users data in their descriptions.}}
	\label{table:nopolicy}
\end{table}

Fig.~\ref{noprivacy} in Appendix illustrates an example where the skill "Name My Grandkids" includes in its description that it asks the users for personal information and stores it for future use. In another case, the skill "Lapel Athletics" requires the device location according to its description. But both these skills do not provide a privacy policy. Table~\ref{table:nopolicy} lists skills which are supposed to have a privacy policy but do not provide one.

\subsubsection{Cross-platform inconsistency}
For a few voice-apps that are present on both Alexa and Google platforms, we found that the privacy policies provided with each are not the same.
Comparing Google actions and Alexa skills, we found that 23 voice-apps which are present on both the platforms have differences in the privacy policy links provided despite the name, the descriptions and the developer name being the same. 13 of these pairs have different privacy policies links all together. For example, the skill "Did Thanos Kill Me" uses a duplicate privacy policy link \url{"https://getstoryline.com/public/privacy.html"} (shown in Table~\ref{table:duplicate}), but the corresponding Google action version provides a specific privacy policy.
Since Google requires every action to provide a privacy policy link, developers provide one with the Google action but may choose to not provide one along with the Alexa skill since the Alexa platform doesn't have this requirement. We found 10 such pairs of skills. 
Table~\ref{appendix:table:cross-platform} shows the list of these skills with cross-platform inconsistency.
\begin{table}[h!]
	\vspace{-7pt}
	\centering
	\resizebox{7.5cm}{!}{
		\begin{tabular}{  >{\centering\arraybackslash}m{3cm} >{\centering\arraybackslash}m{6.5cm} }
		\toprule[1.5pt]
		Issue & Skill name \\
		\hline	
		{\bf Different privacy policies provided in a skill \& action pair} 	& 	VB Connect, Orbit B-Hyve, Freeze Dancers, Burbank Town Center, New York Daily News, uHoo, RadioMv, MyHealthyDay, Real Simple Tips, Did Thanos Kill Me, Central Mall Lawton, Creative Director, The Hartford Small Business Insurance\\
		\hline

	{\bf A skill doesn't have
	a privacy policy while the Google action version has one} & Triple M NRL, Collective Noun, Sleep by Nature Made, Coin Control, Those Two Girls, Radio Chaser, A Precious Day, Running facts, Ash London Live, Weekend Breakfast\\
			\bottomrule[1.5pt]
		\end{tabular} 
	}
	\vspace{2pt}
	\caption{Same voice-apps with different privacy policies on two VA platforms.}\label{appendix:table:cross-platform}
\end{table}

\subsubsection{Potential noncompliance with legal regulations}

We observed skills that collect personal information being published on the Amazon Alexa skills store under the kids category without providing a privacy policy. For example, Table~\ref{table:nopolicy} lists 3 skills (which are marked with "Kids" in the table) in the kids category lacking a privacy policy. This is not compliant with the COPPA regulations which require every developer collecting personal information from children to follow certain rules. Providing a privacy policy with accurate information about the data being collected and what it is used for is one of the main requirements. The objective is to clearly let the parents know about what personal information can be collected by the skill from their children.
Health related information can also be collected by a skill through the conversational interface without providing a privacy policy even though only the user can decide whether to provide it or not. But still having the capability to do so might be a violation of the HIPAA (Health Insurance Portability and Accountability Act) regulation.
CalOPPA (California Online Privacy Protection Act) requires developers to provide a privacy policy that states exactly what data can be collected from users. In Sec.~\ref{sec:Finding2:general}, we found that 137 kids skills provide general information without providing specifics on what personal data they actually collect. These voice-apps and their privacy policies may not be in compliance with legal regulations.

\subsection{Usability issues}\label{sec:Finding3}
\subsubsection{Lengthy privacy policies} 

One of the main problems associated with privacy policies regardless of the type of service it is provided with is the length of the privacy policy document. Most developers write long policies that decreases the interest that a user has in reading it. From the analysis of the privacy policies in our datasets, as shown in Fig.~\ref{cdf:wordlength}, we observed that 58\% of the privacy policies have more than 1,500 words. Being a legal document, it takes an average of 12 mins to read 1,500 words. This makes the privacy policy hard to read for the users and almost impossible to be read out through voice. The privacy policies of Google and Amazon themselves are more than 4,300 words each. The average number of words in a privacy policy provided along with Alexa skills is 2,336 and that of Google actions is 1,479. This results in users frequently skipping reading the privacy policy even if it is provided. The participants of the user study we conducted as shown in Sec.~\ref{userstudy} complained about the length of the privacy policy being the major reason for them not reading the privacy policy.   

\subsubsection{Hard to access}

The constrained interfaces on VA devices pose challenges on
effective privacy notices.  
According to the current architecture of Amazon Alexa and the Google Assistant, the privacy policy is not available directly through VA devices used at home like the Amazon Echo and the Google Home. No prompt is delivered either during any part of the users interaction with the voice-app that requests the user to take a look at the privacy policy. If at all the user wants to view the privacy policy, he/she has to either find the voice-app listing on the store webpage or check the companion app using smartphones, and find a voice-app's privacy policy URL provided in the listing. The permissions set by the developer to collect personal information from users is shown as a prompt in the smartphone companion app to the user while enabling the voice-app. But as mentioned in the Sec.~\ref{missingpolicy}, developers do not necessarily have to take permission from user and can instead collect it during the conversation. 
We discuss solutions to improve the usability of privacy notice for voice-apps in Sec.~\ref{intent}.

%% file: usersurvey.tex
\section{User Study}\label{userstudy}

We conducted an online user study using the Amazon Mechanical Turk crowdsourcing platform~\cite{AmazonTurk}, and our study has received an IRB approval. Different from prior user studies that focus on understanding security and privacy concerns of VA devices~\cite{Zeng:2017:SOUPS, Maurice:2017:Law, Nathaniel:CHI:2018, Malkin:2019:Privacy, SOUPS:2019}, we aimed to understand how users engage with privacy policies and their perspectives on them. We looked for the frequency of checking the privacy policies and any issues the users might have encountered with it. Our participants were MTurk workers who reside in USA, having a HIT (Human Intelligence Tasks) acceptance rate greater than 98 and have at least 500 HITs approved prior to this study. These filters were added to reduce the amount of junk data that we may have collected. All participants were initially presented with a consent form approved by the IRB office. Participants who did not consent to the form were denied to proceed with the study. We rewarded \$0.2 to each participant who completed the study.  

We had a total of 98 participants who took part in our study. We had included a question to ensure that the user is answering the survey authentically. Based on the responses to this question, we rejected the answers of 7 participants. Our results for the user study were thus based on responses from 91 users. The participants are either Amazon Alexa users or Google Assistant users. We didn’t include participants who use other assistants like Siri and Cortana in our study. We had 66 participants who are Alexa users and 25 participants who use Google assistant at home.

\begin{table}[!h]
	\resizebox{8.5cm}{!}{
		\begin{tabular}{|l|l|l|}
			\hline
			\rowcolor[HTML]{C0C0C0} 
			\textbf{Question}  & \textbf{Response} & \textbf{\% of users} \\ \hline
			& Yes               &  48                   \\ 
			\cline{2-3}
			\multirow{-2}{*}{\begin{tabular}[c]{@{}l@{}}Are you aware of the privacy policies of your  skills/actions?\end{tabular}}                                             & No                & 52                   \\ \hline
			\rowcolor[HTML]{EFEFEF} 
			\cellcolor[HTML]{EFEFEF}                                                                                                                                                           & Rarely         & 73                   \\ \cline{2-3} 
			\rowcolor[HTML]{EFEFEF} 
			\cellcolor[HTML]{EFEFEF}                                                                                                                                                           & Half the time     & 11                   \\ \cline{2-3} 
			\rowcolor[HTML]{EFEFEF} 
			\multirow{-3}{*}{\cellcolor[HTML]{EFEFEF}\begin{tabular}[c]{@{}l@{}}How often do you read the privacy policy of a skill/action?\end{tabular}}                       & Most of the time  & 16                   \\ \hline
			& No                & 66                   \\ \cline{2-3} 
			\multirow{-2}{*}{\begin{tabular}[c]{@{}l@{}}Do you read the privacy policy from the skill/action's \\ webpage/Alexa app?\end{tabular}}                                             & Yes               & 34                   \\ \hline
			\rowcolor[HTML]{EFEFEF} 
			\cellcolor[HTML]{EFEFEF}                                                                                                                                                           & No               & 47                   \\ \cline{2-3} 
			\rowcolor[HTML]{EFEFEF} 
			\cellcolor[HTML]{EFEFEF}                                                                                                                                                           & Maybe             & 21                   \\ \cline{2-3} 
			\rowcolor[HTML]{EFEFEF} 
			\multirow{-3}{*}{\cellcolor[HTML]{EFEFEF}\begin{tabular}[c]{@{}l@{}}Do you know what personal data the skills/actions \\ you use are capable of collecting from you?\end{tabular}} & Yes                & 32                   \\ \hline
			& No                & 79                   \\ \cline{2-3} 
			& Maybe             & 7                    \\ \cline{2-3} 
			\multirow{-3}{*}{\begin{tabular}[c]{@{}l@{}}Do you read the privacy policy before using a new \\ skill / action?\end{tabular}}                                                     & Yes               & 14                   \\ \hline
			\rowcolor[HTML]{EFEFEF} 
			\cellcolor[HTML]{EFEFEF}                                                                                                                                                           & No                & 75                   \\ \cline{2-3} 
			\rowcolor[HTML]{EFEFEF} 
			\cellcolor[HTML]{EFEFEF}                                                                                                                                                           & Maybe             & 7                    \\ \cline{2-3} 
			\rowcolor[HTML]{EFEFEF} 
			\multirow{-3}{*}{\cellcolor[HTML]{EFEFEF}\begin{tabular}[c]{@{}l@{}}Do you read the privacy policy before enabling a \\ kid's skill / action?\end{tabular}}                        & Yes               & 18                   \\ \hline
		\end{tabular}
	}
	\vspace{2pt}
	\caption{Survey responses.}
	\label{userstudy}\vspace{-20pt}
\end{table}

Table~\ref{userstudy} shows the survey responses. When asked about whether they are aware of the privacy policies of the voice-apps they use, about 48\% of the participants claimed that they are aware of it. But when asked about how often they actually read the privacy policy provided by the developer, 73\% responded with "rarely". 11\% responded that they read it half the time. 34\% of our participants said that they use the smartphone app or the skills webpage to read the privacy policy while the rest 66\% said that they never read it. 47\% were not aware of what data is being collected by the skill from them and another 21\% were not entirely sure either. This shows a major usability issue where the users ignore the privacy policy even when it is provided by the developer. When asked about the issues they face with privacy policies, 20\% of the participants responded by saying it is hard to access. 44\% of participants felt that the document was too long. 24\% claimed that they felt inconsistencies between the privacy policy and the skill’s actual functionality and description. Users also had problem with developers not providing a privacy policy at all and the ones provided being not informative. The document being too legal and hard to comprehend was a concern for the users. Only 14\% of participants felt that they always check the privacy policy before enabling a skill. 79\% of our participants did not check the privacy policy before enabling a general skill and 75\% did not check it before enabling a kids skill. This lack of usage of the privacy policy by the users shows the need of the voice assistant platforms to address the concerns and take measures to improve the quality as well as the usability of the privacy policies provided by the developers. We have included a few responses from the participants about their perspectives on whether privacy policies should be required for every voice-app in Table~\ref{table:user-view-privacy-policy} in~{Appendix}.

%% file: discussion.tex
\section{Discussion}\label{Discussion}
In this section, we discuss the limitation of this work and further research that can help in addressing the user frustration over
privacy policies in VA platforms. 

\subsection{Limitation}

We are unable to examine the actual source code of voice-apps. The availability of the source code can largely increase the knowledge of what personal data the voice-app is able to collect and where it is stored. This can be compared with the privacy policy to ensure the developer is not performing any malicious activity or misusing the user's trust. With having no baseline, a future research effort that can be done on this regard is to dynamically test voice-apps by enabling them and check their data collection practices. Most developers provide short descriptions which will introduce the skill/action to end users, but data practices are not frequently defined in the descriptions. Since the data related to voice-apps is very limited, we largely depend on the descriptions provided with the voice-apps. This makes our findings on the inconsistency checking not complete. As mentioned in Sec.~\ref{datapractices}, our focus is on revealing the existence of problematic privacy policies, rather than identifying all the inconsistent privacy policies. For capturing data practices, we use a keyword-based method, and compare the similarity of a privacy policy with our keyword dictionary. However, the keyword set can be incomplete. In our future work, we plan to use machine learning techniques to train a model to identify data practices from natural language documents. Nevertheless, we have collected strong evidence in revealing issues over privacy policies on VA platforms.
In addition, the dataset of Google actions that we collected and used for our study is not complete and does not contain all the voice-apps available in the app store. The actions are listed in pages that automatically load more data when the user reaches the end of the page. Since more and more actions keep getting loaded dynamically, we were unable to use the crawler to automatically get information about all the actions.




\subsection{Why poor-quality privacy policies?}

Amazon Alexa and Google Assistant not explicitly requiring app-specific privacy policies results in developers providing the same document that explains data practices of all their services. This leads to uncertainties and confusion among end users. There are skills with privacy policies containing up to 428 data practices and most of these data practices are not relevant to the skill. Thus these documents do not give a proper understanding of the capabilities of the skill to end users.
The poor quality of privacy policies provided with voice-apps is partially due to the lack of an app-specific privacy policy and due to the lenient certification system. 
During the certification process, the content of a privacy policy is not checked thoroughly when the skill is submitted for certification, which has resulted in a large amount of inactive and broken links and also privacy policies not related to the skill. Some privacy policies mention data practices that are in violation of the privacy requirements that Amazon and Google have set but these voice-apps are still certified.

In some cases, even if the developer writes the privacy policy with proper intention and care, there can be some discrepancies between the policy and the actual code. Updates made to the skill might not be reflected in the privacy policy. This is especially possible with the current VA architecture because the backend code of the skill can be updated at any time by the developer and does not require any re-certification to be made available to the end users. The outdated policy may lead to the developers unintentionally collecting personal information without informing the users.







%% file: privacyintent.tex
\subsection{Privacy policy through voice}\label{intent}

The unavailability of privacy policies through voice requires users to access them over the web or through the apps on their phones. One possible reason for this can be due to the large size of the privacy policies and the time required to read out the long document. Users who only use voice assistant services through their VA devices, may not necessarily be aware of the existence of the privacy policies in the respective stores. Also, it is completely left to the user to decide whether to view the privacy policy or not. There is no approval asked prior to enabling the voice-app for the user. In order to address these issues, we propose to introduce a built-in intent (\ie, functionality) for a voice-app that gives information to users about the privacy policy of the voice-app through a voice response. The major challenge for this is that the privacy policies are usually too long to be read out to users. Thus, the response provided by the built-in intent has to be marginally short.

Prior work has been done to summarize the privacy policies to make it more readable to the user. Tools like Polisis~\cite{DBLP:Hamza:Polisis} and Privacycheck~\cite{privacycheck} conduct privacy policy analysis and represent the data practices mentioned in the document in a simpler form to users. But from our analysis of the skills/actions available in the stores, we have noticed that most privacy policies are general policies and do not necessarily define what the behavior of the voice-app in particular is. 
Since personal information can be collected through the conversational interface, our approach aims in understanding this capability from the voice-app's source code, automatically generating
an easy-to-digest privacy notice, and letting the user know about it through the voice channel. 
To achieve this, we describe our preliminary approach based on the Alexa platform. We take the interaction model of a skill, which is a JSON (JavaScript Object Notation) file and scan for all the slots and their slot types specified. We categorise the built-in slot types based on what type of personal information they can collect. For custom slot types, we compare the values provided with the entries in datasets we assembled of possible values and check for a match. After we get all the types of information that can be collected by the skill, we create a response notifying the user that the skill has these capabilities and advise users to look at the detailed privacy policy provided by the developers.
The intent can be invoked when the skill is first enabled. On opening the skill for the first time,
this brief privacy notice can be read out to the user. This will give the user a better understanding of what the skill he/she just enabled is capable of collecting and using. The users can also ask to invoke this intent later to get a brief version of the privacy policy. 
As our future work, we plan to extend this approach to help developers automatically generate privacy policies for their voice-apps.




%% file: relatework.tex
\section{Related Work}\label{relatedwork}



%
\subsection{Privacy concerns for voice assistants} 


Many research efforts have been undertaken to study user concerns (human factors) about the security/privacy of VA devices~\cite{Zeng:2017:SOUPS, Chung:2017:Computer, Maurice:2017:Law, Nathaniel:CHI:2018, Geeng:2019:WCI, MCLEAN201928, Ammari:2019:MSI, Alexander:2019, Malkin:2019:Privacy, SOUPS:2019}. Fruchter~\etal~\cite{Nathaniel:CHI:2018} used natural language processing to identify privacy
and security related reviews about VA devices from four major online retailers: Target, Walmart, Amazon, and Best Buy. The authors highlighted that users worried about the lack of clarity about the scope of data collection by their voice assistants. 
Through a semi-structured interviews with 17 VA users, Abdi~\etal~\cite{SOUPS:2019} uncovered the lack of trust users have with some of VA use cases such as shopping, and a very limited conception of VA ecosystem and related data activities. 
Malkin~\etal~\cite{Malkin:2019:Privacy} surveyed 116 VA owners and found that half did not know that their recordings were being stored by the device manufacturers. Similarly, authors in~\cite{Zeng:2017:SOUPS, Serena:2018:HCI} conducted interviews on smart home owners to examine user mental models and understand their privacy perceptions of IoT devices. Geeng~\etal~\cite{Geeng:2019:WCI} investigated tensions and challenges that arise among multiple users in smart home environment. Lau~\etal~\cite{Lau:2018:AYL} conducted interviews with both VA users and non-users, and revealed that privacy concerns could be the main deterring factor for new users.

There has been an increasing amount of research on technical attack vectors against VA systems and the corresponding defenses. One line of research is to exploit interpretation errors of user commands by speech recognition, such as voice squatting attack~\cite{Security:2018:Squatting, DangerousSkills:2019}, and generate hidden/inaudible voice commands~\cite{Vaidya:WOOT:2015, Carlini:Security:2016, Zhang:2017:DIV, Roy:NSDI:2018, Yuan:2018:CSA, Schoenherr:NDSS:2019, Yan:2020:Surfingattack}. 
Another line of research focuses on defense mechanisms, including, continuous authentication~\cite{Feng:2017:CAV}, canceling unwanted baseband signals~\cite{Zhang:2017:DIV}, correlating magnetic changes with voice commands~\cite{Chen:2017:ICDCS}, and user presence-based access control~\cite{Lei:CNS:2018}. Our work differs from these previous work in that we investigate the effectiveness of privacy policies provided by voice-app developers.  





\subsection{Privacy policy analysis for mobile apps}

Privacy policies disclose an organization's or developer's data practices. Though researchers have conducted privacy policy analysis on Android platform~\cite{Slavin:ICSE:2016, Yu:2016:DSN, zimmeckEtAlCompliance2017,Wang:2018:ICSE, Zimmeck:PETS:2019, SEC:Benjamin:2019}, there is an absence of privacy policy analysis on VA platforms.
Zimmeck~\etal~\cite{zimmeckEtAlCompliance2017} presented a privacy analysis system for Android to analyze apps' potential non-compliance with privacy requirements, and inconsistencies between privacy policies and apps. Results show that 71\% of apps that lack a privacy policy should have one, and a substantial portion of apps exhibit potential privacy requirement inconsistencies. Wang~\etal~\cite{Wang:2018:ICSE} developed a hierarchical mapping based approach for privacy policy analysis which is able to handle the data inputted by users in addition to the data accessed directly through the mobile device. The user input data is checked for possible privacy leaks and this is used to determine whether the app's privacy policy is in contradiction with this leakage. The consistency between the data collected by the app and the privacy policy provided is verified by using a data flow analysis. 
A major difference of our work from these works is that we rely on voice-app's description to detect inconsistency in privacy policies due to the unavailability of voice-app's source code. To the best of our knowledge, this is the first work to systematically measure the effectiveness
of privacy policies for voice-apps.




%% file: concludion.tex
\section{Conclusion}
In this work, we conducted a comprehensive empirical analysis on privacy policy of 64,720 Amazon Alexa skills and 2,201 Google Assistant actions. We designed an NLP-based approach to capture data practices in privacy policies and descriptions of voice-apps. Our results showed that a substantial number of problematic privacy policies exist in Amazon Alexa and Google Assistant platforms, a worrisome reality of privacy policies on VA platforms.
Google and Amazon even have official voice-apps violating their own requirements regarding the privacy policy. We also conducted a user study to understand users’ perspectives on voice-apps' privacy policies, which reflects real-world user frustrations on this issue. We also discussed possible approaches to improve the usability of privacy policies on VA platforms.

%% file: appendix.tex
\newpage
\begin{appendices}\label{appendices}

\appendixpage
\begin{table}[h]	
	\centering
	\resizebox{7.0cm}{!}{
		{
			\begin{tabular}{  >{\centering\arraybackslash}m{3.5cm}   >{\centering\arraybackslash}m{2cm} >{\centering\arraybackslash}m{2cm} }
				\hline
			\rowcolor[HTML]{EFEFEF}	Category & Skills we crawled & Skills with a privacy policy \\
				\hline
				Business \& Finance  & 3,599 & 1,420  \\  
				\hline
				Connected Car  & 140 & 100 \\  
				\hline	
				Education \& Reference & 7,990 & 1,460 \\  
				\hline				
				Food \& Drink  & 1,379 & 407 \\  
				\hline				
				Games \& Trivia & 11,822 & 1,461\\  
				\hline				
				Health \& Fitness & 2,026 & 844\\  
				\hline				
				Kids& 3,252 & 461\\  
				\hline				
				Lifestyle & 11,080 & 2,693 \\  
				\hline				
				Local  & 1,283 & 377\\  
				\hline				
				Movies \& TV  & 915 & 153\\  
				\hline				
				Music \& Audio  & 9,216 & 3,155\\  
				\hline				
				News  & 6,810 & 2,907  \\  
				\hline				
				Novelty\& Humor & 3,418 & 394 \\  
				\hline				
				Productivity & 4,263 & 1,434\\  
				\hline				
				Shopping  & 342 & 204 \\  
				\hline				
				Smart Home & 2,432 & 2,254 \\  
				\hline				
				Social  & 1,479 & 531 \\  
				\hline
				Sports  & 1,592 & 343 \\  
				\hline
				Travel \& Transportation & 1,187 & 205\\  
				\hline
				Utilities & 1,025 & 191\\  
				\hline
				Weather  & 853 & 150\\  
				\hline
				Total skills & 76,103 & 21,144 \\  
				\hline
			\rowcolor[HTML]{EFEFEF}	Total unique skills & 64,720 & 17,952 \\  
				\hline
			\end{tabular}
		}
	}
	\vspace{-0pt}
	\caption{Alexa skills by category in our dataset. Some skills are classified and listed in multiple categories. After removing the cross-listed duplicates, we obtained 64,720 unique skills, and 17,952 of these skills provide privacy policy links.}
	\label{table:skilldataset}
\end{table}

\begin{table}[h]	
	\centering
	\resizebox{7.0cm}{!}{
		{
			\begin{tabular}{  >{\centering\arraybackslash}m{3.5cm}  >{\centering\arraybackslash}m{2cm}  >{\centering\arraybackslash}m{2cm}}
				\hline
				\rowcolor[HTML]{EFEFEF}	Category & Actions we crawled & Actions with a privacy policy \\
				\hline
				Arts \& Lifestyle & 96 & 85  \\  
				\hline
				Business \& Finance & 159 & 156 \\  
				\hline	
				Communication \& Social & 27 & 11 \\  
				\hline				
				Education \& Reference & 153 & 145 \\  
				\hline				
				Food \& Drink & 152 & 142\\  
				\hline				
				Games \& Fun & 174 & 167 \\  
				\hline				
				Health \& Fitness & 122 & 119\\  
				\hline				
				Kids \& Family & 102 & 101\\  
				\hline				
				Local & 65 & 61\\  
				\hline				
				Movies, Photos \& TV & 52 & 44\\  
				\hline				
				Music \& Audio & 132 & 121\\  
				\hline				
				News \& Magazines & 176 & 80  \\  
				\hline				
				Productivity & 52 & 41\\  
				\hline				
				Shopping & 89 & 86 \\  
				\hline				
				Smart Home & 290 & 274 \\  
				\hline
				Sports & 161 & 142 \\  
				\hline
				Travel \& Transportation & 142  & 136\\  
				\hline
				Weather & 57 & 56\\  
				\hline
				\rowcolor[HTML]{EFEFEF}	 Total actions & 2,201 & 1,967 \\  
				\hline				
			\end{tabular}
		}
	}
	\vspace{-0pt}
	\caption{Google actions by category in our dataset.}
	\label{table:actiondataset}
\end{table}

\begin{table}[h]
	\centering
	\resizebox{8.5cm}{!}{
		\begin{tabular}{  >{\centering\arraybackslash}m{8.5cm} } \Xhline{3\arrayrulewidth}
			Arbonne My Office, Arm My Guardzilla, Ash Timber Flooring, Atrium Health, Best Roomies, Cake Walk, Call Handler, CitySpark Events, Conway Daily Sun Calendar, Crush Calculator, Find My Phone, Flu Season, FortiRecorder, garage control, GINA Talk, group messenger, Hal9000, Happy birthday, Home Workout Exercise Video Fitness 7 Day Videos, hugOne, ISS: Distance From Me?, K5, Kamakshi Cloud's GPS Finder, Kotipizza, Laconia Daily Sun Calendar, Mailbox Assistant, Maui Time Calendar, My Air Quality, My Extra Brain, My Two Cents, Natural Hazards, Neighbor Knocker, Novant Health, OMS Customer Care, Portland Phoenix Calendar, Prayer Time, Record-Journal - Things to Do Calendar, Running Clothes, Running Outfit Advisor, SkyHome, SkyView Academy, The Transit Oracle (Bus Predictions for SF Muni), Thought Leaders, Ticket Finder, Trip Tracker, Trivia Quest, walk cake, What Should I Wear, what's nearby, WP6\\
			\Xhline{3\arrayrulewidth}
		\end{tabular} 
	}
	\caption{Amazon Alexa skills with inconsistency between the privacy policy and description.}\label{table:incompletepolicy}
\end{table}

\vspace{-0pt}
\begin{figure} [!h]
	\begin{center}
		\includegraphics[width=0.47\textwidth]{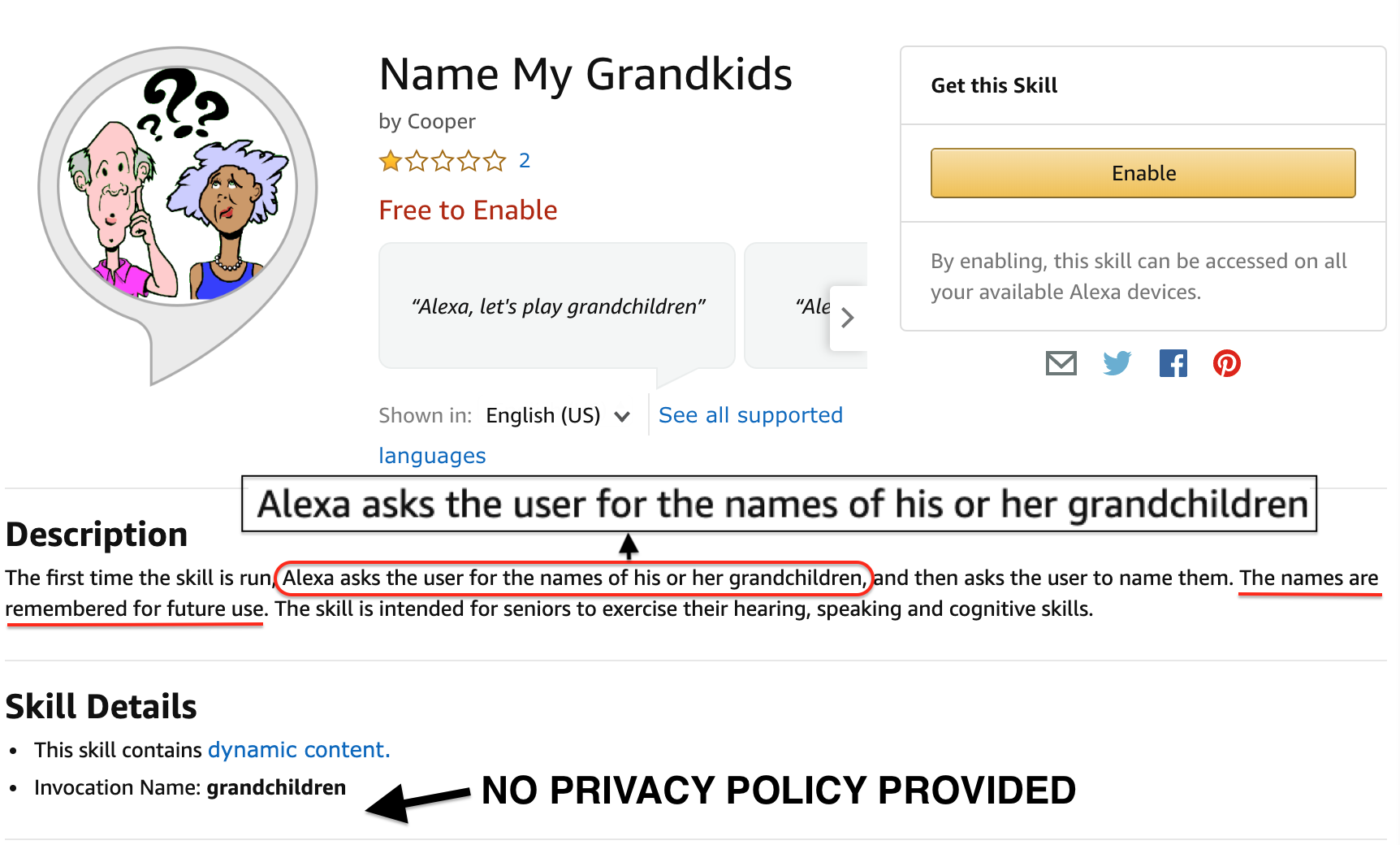}
	\end{center} 
	\vspace{-10pt}
	\caption{Although the skill description mentions collection of personal information, no privacy policy is provided. }
	\label{noprivacy}\vspace{10pt}
\end{figure}

\begin{table}[h]	
	\centering
	\resizebox{8.5cm}{!}{
		{
			\begin{tabular}{ | >{\centering\arraybackslash}m{1.5cm} | >{\centering\arraybackslash}m{9cm}|}
				\hline
				 {\bf Question} & {\bf Do you think all skills should have a privacy policy?}  \\
				\hline
				\multirow{16}{*}{ {\bf Responses}}
				& A privacy policy is always necessary to give users a piece of mind.\\
				 \cline{2-2}
				& Users should be able to know the risks involved such as if others could be listening in illegally.\\
				\cline{2-2}
				& Privacy policy is definitely required so it can assure consumers that it is unlikely that malicious actions will occur with their data.\\
				\cline{2-2}
				& They should be made easily accessible too.\\
				\cline{2-2}
				& Required if in fact there are things that customers should be warned about prior to using it.\\
				\cline{2-2}
				& A privacy policy would be completely necessary. I feel like the skills need to disclose everything being done with a user's data.\\
				\cline{2-2}
				& But it should be easily explained and controls easily learned.\\
			\cline{2-2}
				& If data is being collected, this is personal information that the user should have some control over.\\
				\cline{2-2}
				& It needs to be more digestible so people will actually read it.\\
				\cline{2-2}
				& I do think it is necessary to have a privacy policy, but I do think it should be short and easy to understand.\\
				\hline				
			\end{tabular}
		}
	}
	\vspace{-0pt}
	\caption{{User's view on the necessity of privacy policies. We present a few selected responses received from the participants in our user study when asked the question "Do you think all skills should have a privacy policy?"}}
	\label{table:user-view-privacy-policy}
\end{table}

\end{appendices}
